\documentclass[11pt,a4paper]{article}
\pdfoutput=1

\usepackage[latin1]{inputenc}
\usepackage[english]{babel}
\usepackage{myway}

\hypersetup{pdftitle={Fluxbranes: Moduli Stabilisation and Inflation}, 
pdfauthor={Arthur Hebecker, Sebastian Kraus, Dieter L{\"u}st, 
Moritz K\"untzler, Timo Weigand}, 
pdfsubject={String theory model building},
bookmarksopen, bookmarksnumbered, bookmarksopenlevel=2}

\newcommand{\beq}{\begin{equation}} 
\newcommand{\eeq}{\end{equation}}
\newcommand{\bal}{\begin{aligned}}  
\newcommand{\eal}{\end{aligned}}
\newcommand{\bea}{\begin{eqnarray}} 
\newcommand{\eea}{\end{eqnarray}}

\newcommand{\ccA}{\mathcal{A}}
\newcommand{\ccD}{\mathcal{D}}

\newcommand{\ccE}{\mathcal{E}}
\newcommand{\ccF}{\mathcal{F}}
\newcommand{\ccB}{\mathcal{B}}
\newcommand{\ccV}{\mathcal{V}}
\newcommand{\ccK}{\mathcal{K}}
\newcommand{\ccC}{\mathcal{C}}

\newcommand{\ccO}{\mathcal{O}}

\newcommand{\ccL}{\mathcal{L}}

\newcommand{\mkk}{m_{\mbox{\tiny KK}}}

\newcommand{\secref}[1]{section~\ref{#1}}
\newcommand{\Secref}[1]{Section~\ref{#1}}

\newcommand{\appref}[1]{appendix~\ref{#1}}

\title{Fluxbranes: Moduli Stabilisation and Inflation}

\preprint{MPP-2012-113\\
LMU-ASC 47/12\\
}

\author[1]{Arthur Hebecker\note[ ]
{\href{mailto:A.Hebecker@ThPhys.Uni-Heidelberg.de}
{A.Hebecker@ThPhys.Uni-Heidelberg.de}},}
\author[1]{Sebastian C. Kraus\note[ ]
{\href{mailto:S.Kraus@ThPhys.Uni-Heidelberg.de}
{S.Kraus@ThPhys.Uni-Heidelberg.de}},}
\author[2]{Moritz K\"untzler\note[ ]
{\href{mailto:moritz.kuentzler@kcl.ac.uk}
{Moritz.Kuentzler@kcl.ac.uk}},}
\author[3]{Dieter L{\"u}st\note[ ]
{\href{mailto:Dieter.Luest@lmu.de}{Dieter.Luest@lmu.de}},}
\author[1]{\\ and Timo Weigand\note[ ]
{\href{mailto:T.Weigand@ThPhys.Uni-Heidelberg.de}
{T.Weigand@ThPhys.Uni-Heidelberg.de}}}
\affiliation[1]{Institut f\"ur Theoretische Physik, Universit\"at Heidelberg, 
Philosophenweg 19, D-69120 Heidelberg\vspace{0.1cm}}
\affiliation[2]{Department of Mathematics, King's College, London, The Strand, WC2R 2LS, London, \\
United Kingdom   \vspace{0.1cm}}

\affiliation[3]{Arnold-Sommerfeld-Center, Ludwig-Maximilians-Universit\"at, 
Theresienstrasse 33, \\  D-80333 M{\"u}nchen\vspace{0.1cm}}
\affiliation[3]{Max-Planck-Institut f\"ur Physik, F\"ohringer Ring 6, 
D-80805 M\"unchen\vspace{0.1cm}}

\abstract{
Fluxbrane inflation is a stringy version of $D$-term inflation in which two 
fluxed D7-branes move towards each other until their (relative) gauge flux 
annihilates. Compared to brane-antibrane inflation, the leading-order 
inflationary potential of this scenario is much flatter. In the present paper we first discuss a new explicit moduli stabilisation procedure combining the $F$- and $D$-term scalar potentials: It is based on fluxed D7-branes in a geometry with three large four-cycles of hierarchically different volumes.
Subsequently, we combine this moduli stabilisation with the fluxbrane inflation idea, demonstrating in particular that CMB data 
(including cosmic string constraints) can be explained within our setup of \textit{hierarchical} large volume CY compactifications. We also indicate how the $\eta$-problem is expected to re-emerge through higher-order corrections and how it might be overcome by further refinements of our model. Finally, we explain why recently raised concerns about constant FI terms 
do not affect the consistent, string-derived variant of $D$-term inflation discussed in this paper.}

\begin{document}

\noindent January 24, 2013

\vspace*{-0.8 cm}
\maketitle

\section{Introduction}

Realising inflation in string theory
has turned out to be a challenging problem. Amongst the models investigated in this context, many fall into a popular class known as 
brane inflation \cite{Dvali:1998pa}. Here the inflaton is associated with the relative distance between a brane and an antibrane \cite{Burgess:2001fx} or between two D-branes \cite{GarciaBellido:2001ky,Dasgupta:2002ew}.

This is a rather attractive setting because, in analogy to $D$-term \cite{Binetruy:1996xj,Halyo:1996pp} or (more generally) hybrid inflation \cite{Linde:1991km,Linde:1993cn}, the energy density is dominated by 
a constant term as long as the branes are far apart. In this regime, 
the potential can be naturally flat. Later, once the branes have approached each other up to a certain critical distance, tachyon condensation takes the potential to zero almost instantaneously.
 
Unfortunately the simplest variant, where a brane and an antibrane 
annihilate at the end of inflation, cannot work since the size of the compact space does not allow for a sufficient brane-antibrane separation
\cite{Burgess:2001fx}. In other words, in spite of the promising idea, 
the potential turns out to be too steep. While this can 
in principle be overcome by strong warping \cite{Kachru:2003sx},
one is eventually forced to play various higher-order corrections off against each 
other \cite{Baumann:2006th,Baumann:2007np,Krause:2007jk,Baumann:2007ah}. This amounts to fine-tuning the inflaton potential such that inflation occurs close to an inflection point. One might thus feel that the original idea of `stringy hybrid inflation' is lost.

To maintain the `hybrid inflation paradigm', we therefore return to 
brane inflation models with two D-branes \cite{GarciaBellido:2001ky,Dasgupta:2002ew}. The first of these is 
`inflation from branes at angles' \cite{GarciaBellido:2001ky}, which has its natural home in Type IIA orientifold models with D6-branes. Our fluxbrane proposal, to be discussed momentarily, can be viewed as the Type IIB mirror dual of this class of models. It hence has the enormous advantage of better control concerning moduli stabilisation. The second, D3/D7 inflation
\cite{Dasgupta:2002ew}, has serious issues which we aim to overcome:
In D3/D7 inflation, supersymmetry is broken by non-selfdual flux on the D7-brane. As the D7-brane is much heavier than the D3-brane it is, in general, the latter which moves in the geometry of the fluxed D7-brane. As it turns out, this realisation of D3/D7 inflation suffers from a similar problem as the brane-antibrane proposal \cite{Burgess:2001fx}, i.e.\ the potential is too steep. As possible ways out, a fine-tuned `small field' version or the use of a highly anisotropic orbifold have been suggested in \cite{Haack:2008yb}. Alternatively, one might consider a setting where the fluxed D7-brane moves, probing the geometry produced by a large-$N$ stack of D3-branes \cite{Hsu:2003cy}. 

While these suggestions certainly warrant further detailed investigation, we 
believe that ``fluxbrane inflation'' \cite{Hebecker:2011hk} provides a more direct approach to `stringy $D$-term inflation'. The basic idea behind fluxbrane inflation is easy to state: The inflaton is associated with the relative distance of two D7-branes which carry non-selfdual worldvolume flux $F$. Once the branes come sufficiently close, (part of) the flux annihilates and the universe reheats. While this tachyon condensation process releases an energy density $\sim F^2$ (the constant term in the potential), the brane-distance-dependent attractive part of the potential is $\sim F^4$ (cf.\ section 1 of \cite{Hebecker:2011hk} for an intuitive explanation of this important technical fact). In other words, very schematically the inflaton potential has the form
\begin{equation}
 V(\varphi) \sim F^2 + F^4 \log \varphi \, .
\end{equation}
In the large-volume limit, where the flux density $F\to 0$, this gives us an obvious 
advantage over brane-antibrane scenarios: The $\varphi$-dependent term is 
more strongly suppressed than the constant, making the (leading order) potential flat enough for many e-foldings. Furthermore, when comparing to D3/D7 inflation (which is otherwise closely related), we do not have to 
appeal to the very large $N$ of the D3-brane stack in order to make the D7-brane move. We finally note that, in addition to being T- or 
mirror-dual to \cite{GarciaBellido:2001ky}, a different T-duality relates 
our setting to Wilson line inflation \cite{Avgoustidis:2006zp}.

In the analysis of fluxbrane inflation in \cite{Hebecker:2011hk}, moduli stabilisation was essentially taken for granted. This is a strong
assumption for two reasons. On the one hand, our scenario requires 
specific values for certain parameters of the compactification. It has 
therefore to be checked that these values can indeed be attained. On the 
other hand, the physical effects invoked to stabilise moduli tend to 
destroy the flatness of the inflationary potential, an effect well familiar also in other classes of brane inflation \cite{Kachru:2003sx,Berg:2004ek,Berg:2004sj,McAllister:2005mq,Baumann:2006th}. Hence, the flatness of
the potential has to be checked {\it after} moduli stabilisation. 

As we will demonstrate in some detail in the following, moduli 
stabilisation is possible in the phenomenologically required regime. 
This is the focus of the present analysis. Concerning the flatness of the 
potential after moduli stabilisation, we will at least be able to identify 
the most dangerous higher-order corrections. We will then present a
strategy for suppressing them within our scenario, postponing a more 
thorough discussion to a further publication \cite{wip}.

The size of curvature perturbations in fluxbrane inflation is 
governed by the inverse volume, forcing us into a regime where the volume $\ccV$ is very large. K\"ahler moduli stabilisation is then naturally realised in the 
Large Volume Scenario \cite{Balasubramanian:2005zx,Conlon:2005ki}. The 
latter is based on the interplay between $\alpha'$- and non-perturbative corrections to K\"ahler and superpotential, resulting in a non-supersymmetric AdS vacuum. This vacuum is then uplifted by some additional positive contribution to the vacuum energy density, such as fluxes on D7-branes \cite{Burgess:2003ic,Villadoro:2005yq,Achucarro:2006zf,Haack:2006cy,
Cremades:2007ig,Krippendorf:2009zza,Cicoli:2011yh,Cicoli:2012vw} or $\overline{\text{D3}}$-branes in a warped throat \cite{Kachru:2002gs,Kachru:2003aw,Kachru:2003sx}. As will be worked out in \secref{ModStab}, using the first of these two possibilities in the context of fluxbrane inflation requires two independent flux-effects: One of them annihilates at the end of inflation, when the two relevant D7-branes meet.
The other flux can not annihilate given that certain topological requirements 
are fulfilled. This flux is responsible for the uplifting to a Minkowski vacuum, which has to remain intact after reheating.

Appropriately suppressing the cosmic string contribution to CMB fluctuations is a crucial issue in brane or $D$-term inflation. In our scenario, the stability of cosmic strings is not completely trivial (cf. the discussion 
in \cite{Binetruy:2004hho}). To be on the safe side, we consider the worst-case scenario of topologically stable (local) cosmic strings rather than their semilocal cousins \cite{Vachaspati:1991dz,Hindmarsh:1991jq,Urrestilla:2004eh,Binetruy:2004hho,Dasgupta:2004dw}. Cosmic string suppression then requires a hierarchy of four-cycle volumes in the internal manifold \cite{Hebecker:2011hk}. This forces us to go beyond the simple `warm-up' model of \secref{ModStab}. Thus, in \secref{Anisotropic Setup}, we embed our model of inflation in a {\it hierarchical} Large Volume Scenario, along the lines of \cite{Cicoli:2007xp,Cicoli:2008va}.

In this scenario, one is dealing with at least three K\"ahler moduli and 
stabilisation relies, in addition to  $\alpha'$- and instanton effects, on 
$g_s$- or loop-corrections to the K\"ahler potential \cite{vonGersdorff:2005bf,Berg:2005ja,Berg:2005yu,Berg:2007wt}. In our case it turns out that, for being able to satisfy all phenomenological requirements, the minimal number of K\"ahler moduli is four.
We work out in detail how the three resulting contributions to the scalar potential combine with the $D$-term to give rise to a Minkowski/de Sitter minimum in which 
all K\"ahler moduli are stabilised. 

To the extent that this is possible, we follow the discussion of the simple 
model of \secref{ModStab}. In particular, the K\"ahler modulus of the `small'
four-cycle carrying the instanton can be integrated out right away. We are 
then left with the K\"ahler moduli of the three large four-cycles. It is 
convenient to think in terms of the Calabi-Yau volume and two dimensionless 
ratios of two-cycle volumes. The latter are fixed by the interplay of 
$g_s$-corrections and the $D$-term. Thus, one arrives at a fairly simple 
potential involving only the total volume. It is then possible to 
demonstrate stabilisation maintaining (almost) complete analytical control.

We end up with large overall volume ($\ccV \simeq 3.5\times 10^7$ in units of 
$\ell_s$ in the 10d Einstein frame) and large tree-level superpotential ($W_0 \simeq 2\times 10^3$). The string coupling $g_s$ is well in the perturbative regime ($g_s \simeq 3\times 10^{-2}$). Therefore, moduli stabilisation is achieved in a way consistent with all phenomenological requirements formulated in \cite{Hebecker:2011hk}.

Such a large value of $W_0$ certainly bears the danger of violating the D3 tadpole constraint. In particular, translated into the language of F-theory compactifications, one needs an elliptically fibered fourfold $X_4$ with a large Euler characteristic $\chi(X_4)$ to allow for a sizable $W_0$ \cite{Denef:2004ze}. As we will see, the Euler numbers which have been found in \cite{Klemm:1996ts} are sufficient for our purposes.

Combining the idea of ``fluxbrane inflation'' (with all its phenomenological constraints) with moduli stabilisation within hierarchical Large Volume Scenarios turns out to lead to a rather restrictive setting: For example, it was not possible to find a model with only three K\"ahler moduli in which one has parametric control over the size of the ratio $m_{3/2}/\mkk$ and, at the same time, has all intermediate-size two-cycles parametrically large. While we were able to overcome this issue in a model with four K\"ahler moduli, we had to give up the idea of realising the uplift of the AdS minimum to Minkowski via a $D$-term. Thus, in \secref{Anisotropic Setup} we need to do this uplift by means of a different contribution to the vacuum energy density which, for concreteness, we chose to be $\overline{\text{D3}}$-branes. It would be interesting to investigate in more detail under which circumstances the idea of an uplift induced by fluxes on D7-branes can be realised. However, this is beyond the scope of this paper.

To have a convincing model of stringy $D$-term inflation, it is of course necessary to control inflaton mass corrections from the $F$-term potential. There are numerous sources for such terms. First of all, we have to choose the 
flux such that it does not stabilise the D7-brane position via an explicit appearance in the superpotential. The requirement for this (the flux has to be a two-form obtained by pullback from the ambient space) has already 
been discussed in \cite{Hebecker:2011hk}. Next, it is clear that the 
K\"ahler potential for D7-brane positions is non-minimal. More specifically, 
the D7-brane modulus $\zeta$ appears in the 4d supergravity K\"ahler potential in the form \cite{Jockers:2004yj,Jockers:2005zy}
\begin{equation}
 K \supset - \log (S + \overline{S} - k(\zeta,\overline{\zeta}))\,.
\label{ksz}
\end{equation}
Here $k(\zeta,\overline{\zeta})$ is the K\"ahler potential on the moduli space of D7-brane positions and $S$ is the axio-dilaton. It is thus apparent that, as soon as fluxes are turned on and some non-zero $W_0$ is generated, a mass term for $\zeta$ is induced in the $F$-term scalar potential. This generically leads to an $\eta$-problem even if $W_0$ is $\zeta$-independent. 

An exception occurs if $k$ possesses a shift-symmetry, i.e. $k(\zeta,\overline{\zeta})=k(\zeta + \overline{\zeta})$. This can be the case
in $K3\times K3$ and certain orbifold models (see \cite{Shandera:2004zy,Hsu:2003cy,Hsu:2004hi,McAllister:2005mq} for similar attempts in the context of D3/D7 inflation). Additionally, as we will explain in somewhat more detail in \secref{FlatDir}, we expect that the moduli spaces of generic Calabi-Yau compactifications have corners where such a shift-symmetric structure arises at least approximately.
However, this is not the end of the story: One can see from equation \eqref{ksz}
that the moduli spaces of axio-dilaton $S$ and D7-brane modulus $\zeta$ 
are intertwined in a non-trivial fibration. Hence, the $S$-dependence 
of $W_0$ potentially induces a non-trivial $\zeta$-dependence.

Furthermore, $g_s$-corrections to the K\"ahler potential, which are used in \secref{Anisotropic Setup} to stabilise part of the K\"ahler moduli, are known to depend on open string moduli. All these effects can possibly spoil the nice properties of the fluxbrane inflation scenario. \Secref{FlatDir} is dedicated to investigating these issues. While we are optimistic that a viable region 
in the parameter space can be found, the analysis of this section is not yet conclusive. This set of problems is still under investigation \cite{wip}.

Finally, we want to discuss our model in the light of an old question regarding the nature of the $D$-term in supergravity (see e.g.\ \cite{Witten:1985bz,Binetruy:2004hho}). Recently, this issue has received some renewed interest in the work of
\cite{Komargodski:2009pc,Dienes:2009td,Komargodski:2010rb,Seiberg:2010qd,
Distler:2010zg,Banks:2010zn,Hellerman:2010fv}. In particular, the authors of \cite{Komargodski:2009pc,Komargodski:2010rb} showed that in supergravities which emerge in the low-energy limit of a string compactification it is inconsistent to have a constant FI term. This is in complete agreement with the well-known structure of the $D$-term potential arising from string compactifications. These results also rule out models in which the FI term is dynamical at first, but then assumed to be stabilised at a certain scale such that it can be viewed as a constant from the point of view of a supersymmetric low-energy effective theory.
We address this issue in \secref{consistency} by computing the relevant moduli masses. It turns out that the forbidden regime with an effectively constant 
$D$-term can not be realised in our stringy setting and is indeed not 
necessary for $D$-term inflation to work.

\section{Moduli Stabilisation - Basic Setup}
\label{ModStab}

\subsection{Phenomenological Requirements}
\label{pheno1}
Fluxbrane inflation \cite{Hebecker:2011hk} is a stringy realisation of $D$-term hybrid inflation \cite{Binetruy:1996xj,Halyo:1996pp}. In this scenario two D7-branes in a Type IIB orientifold compactification on a Calabi-Yau threefold ${ X_3}$ are wrapped around two representatives of a holomorphic divisor class of $X_3$ (see \cite{Blumenhagen:2006ci} for a review). The modulus which describes the relative deformation of the two branes is associated with the 4d inflaton. Supersymmetry is broken by gauge flux on the D7-branes. To this end it is most convenient to describe the $U(1)$ gauge theories on the D7-branes in terms of their \textit{overall} ($U(1)_+$) and \textit{relative} ($U(1)_-$) piece. We will have more to say about gauge flux for the $U(1)_+$ theory later on. What is important for the inter-brane potential is the relative gauge flux $\ccF_-$ in terms of which the effective potential for the canonically normalised inflaton $\varphi$ is given by \cite{Hebecker:2011hk}
\begin{equation}\label{genericInflatonPotential}
 V(\varphi) = V_0 \left(1 + \alpha \log \left(\frac{\varphi}{\varphi_0}\right)\right)
\end{equation}
with
\begin{gather}
 V_0 = \frac{1}{2}g_{\rm YM}^2 \xi_-^2\,\,\, , \hspace{1cm} \alpha = \frac{g_{\rm YM}^4}{32\pi^3}\left(\int_{\rm D7} J\wedge \ccF_-\right)^2,\hspace{1cm}
\varphi_0^2\sim \xi_-\,\,,
\\
 g_{\rm YM}^2 = \frac{2\pi}{ \frac{1}{2}\int_{\rm D7} J\wedge J}\,\,, \hspace{1cm} \xi_- = \frac{1}{4\pi}\frac{\int_{\rm D7} J\wedge \ccF_-}{\ccV}\, .
\end{gather}
The $D$-term for $U(1)_-$ is denoted by $\xi_-$. 
In the above formulae it has already been implemented that the D3-brane charge induced by the relative flux vanishes, \mbox{$\int_{\rm D7} \ccF_- \wedge \ccF_- = 0$}. As discussed in \cite{Hebecker:2011hk} this specialisation helps circumvent phenomenological constraints due to cosmic string production at the end of inflation. The reference field value in the logarithm is chosen to be the critical field value $\varphi_0$ below which a tachyon appears in the spectrum.\footnote{A different choice, $\varphi_0 \rightarrow \varphi_0 \times \text{const.}$, would be irrelevant at our level of precision.} Throughout this paper we will measure 4d quantities in units of the reduced Planck mass $M_p$, while internal quantities such as lengths are measured in units of $\ell_s$ in the 10d Einstein frame (cf.\ \appref{Definitions and conventions}).
\par

As discussed in the introduction, the energy-density during inflation comes primarily from the constant term in the potential, while the logarithm presents only a small variation of that constant. When $\varphi$ approaches the critical value $\varphi_0$, tachyon condensation sets in and the universe reheats.

\par

The potential \eqref{genericInflatonPotential} is subject to several phenomenological constraints \cite{Hebecker:2011hk}: First, one can show that the slow-roll conditions are satisfied very naturally in the large volume regime. Secondly, the prediction for the spectral index $n_s$ in terms of the number of e-foldings $N$
\begin{equation}
 n_s \simeq 1-\frac{1}{N} = 0.983 \hspace{0.3cm} \text{for}\hspace{0.3cm} N=60
\end{equation}
agrees with experiment at the level of $2\sigma$ ($n_s = 0.968\pm 0.012$ at $1\sigma$ according to WMAP7 \cite{Komatsu:2010fb}). Finally, the amplitude of adiabatic curvature perturbations $\tilde{\zeta}\equiv V^{3/2}/V'$ is determined by measurements as
\begin{equation}\label{amplitude curvature perturbations}
 \frac{2N}{\tilde{\zeta}^2} = \frac{\alpha}{V_0} = \frac{2\ccV^2}{\frac{1}{2}\int_{\rm D7} J \wedge J} \simeq 4.2 \times 10^8   \hspace{0.3cm} \text{for}\hspace{0.3cm} N=60.
\end{equation}
Assuming that, for the present purposes, the internal manifold can be characterised by a single length scale $R$ we find
\begin{equation}
 R \simeq 11, \hspace{0.3cm} \ccV \simeq 1.7 \times 10^6.
\end{equation}
To the best of our knowledge, the only way to obtain such a large volume in Type IIB string compactifications is in the Large Volume Scenario \cite{Balasubramanian:2005zx,Conlon:2005ki}. In the remainder of this section we therefore outline how moduli stabilisation in fluxbrane inflation can work in principle in this setting.

\subsection{Moduli Stabilisation in the Large Volume Scenario - General Setup}
\label{The LVS}
It was found in \cite{Balasubramanian:2005zx,Conlon:2005ki} that under certain topological conditions there exists a non-super\-symmetric AdS minimum of the scalar potential of Type IIB string theory compactified on a Calabi-Yau orientifold. This minimum appears at an exponentially large volume of the internal manifold and is therefore suitable for our purposes. To find this minimum one applies a two-step procedure: First, the complex structure moduli and the axio-dilaton are stabilised via bulk fluxes \cite{Giddings:2001yu} and integrated out at a high scale, giving rise to a constant tree-level superpotential $W_0$. Due to the `no-scale' structure of the K\"ahler-moduli K\"ahler potential the resulting leading order $F$-term potential is identically zero. A non-zero scalar potential arises at subleading order through $\alpha'$-corrections in the K\"ahler potential and non-perturbative corrections in the superpotential. In a second step one then minimises the effective potential for the K\"ahler moduli resulting from these higher order effects.

D3-instantons can wrap internal four-cycles of the Calabi-Yau manifold. The corrected superpotential is given by
\begin{equation}\label{superpotential corrected}
 W = W_0 + \sum_{p} A_p e^{ - a_p T_p},
\end{equation}
where the $T_p = \tau_p + i b_p$ denote the complexified K\"ahler moduli of the instanton four-cycles. The constants $a_p$ are given by $a_p = 2\pi$, while the Pfaffian prefactor $A_p$ depends on the complex structure moduli and the axio-dilaton (which are assumed to be fixed already) as well as the open string moduli. The latter dependence could well be an issue for the viability of our brane inflation model. For example, it is known that in the presence of D3-branes the one-loop Pfaffian $A_p$ involves the D3-brane position \cite{Baumann:2006th}. A similar effect was argued to occur for D7-branes which carry flux with non-vanishing induced D3-brane charge \cite{Marchesano:2009rz}. Recall from the discussion below (\ref{genericInflatonPotential}) that our flux ${\cal F}_-$ is chosen such that the induced D3-brane charge vanishes (and the same can also be imposed on ${\cal F}_+$). The effect of  \cite{Baumann:2006th,Marchesano:2009rz} is  therefore not expected to occur in our setup. It remains an open question whether, for D7-branes, there is a possible open string dependence of the non-perturbative superpotential beyond these effects. In particular, one must sum over all flux configurations on the D3 instantons  \cite{Grimm:2011dj}, which may introduce such a dependence via the flux induced \mbox{D(-1)} charge. This could be avoided in geometries for which $H^{(1,1)}$ of the instanton divisor only contains elements which are even under the orientifold involution, such that the instanton cannot carry flux  \cite{Grimm:2011dj}. 

The second ingredient apart from the superpotential (\ref{superpotential corrected}) is the K\"ahler potential including $\alpha'$-corrections (which can be shown to be the leading corrections in inverse powers of the volume \cite{Cicoli:2007xp})
\begin{equation}\label{Kahler corrected}
 K = -2\log\left(\ccV + \frac{\xi}{2 g_s^{3/2}}\right)  - \log (S+ \overline{S})+ K_{\text{cs}}.
\end{equation}
Here $\xi = - \frac{\zeta(3)\chi(X_3)}{2(2\pi)^3}$ and $\chi(X_3)$ is the Euler characteristic of the Calabi-Yau threefold $X_3$. Thus the resulting $\alpha'$-contribution to the potential is $\sim 1/\ccV^3$. On the other hand the non-perturbative corrections in the superpotential are exponentially small, leading to a contribution $\sim e^{-a_p\tau_p}/\ccV^2$. For both contributions to be equally relevant for creating a large volume minimum of the scalar potential, one has to require some of the four-cycles on which instantons are wrapped to be exponentially smaller than the overall volume of the threefold. Suppose that there is one such small four-cycle whose modulus we will call $\tau_s$ and whose intersection form is `diagonal' with respect to all other four-cycles in the sense that the only non-vanishing triple intersection number involving $\tau_s$ is its triple self-intersection\footnote{Note that in our conventions $\kappa_{sss}>0$. As the small four-cycle with modulus $\tau_s$ is contractible to a point, this means that in the expansion of the K\"ahler form $J$ the coefficient $t^s$ of the $(1,1)$-form $\omega_s$ is negative, $t^s <0$. Here, $\omega_s$ is Poincar\'{e} dual to the small four-cycle.} $\kappa_{sss}$. Then the volume can be written in terms of the four-cycle moduli as
\begin{equation}\label{generic volume}
 \ccV = \tilde{\ccV}(\tau_{q,q\neq s}) -  c \tau_s^{3/2},
\end{equation}
where $c$ is related to $\kappa_{sss}$ as $c = \frac{2^{3/2}}{3 ! \sqrt{\kappa_{sss}}}$. Furthermore, we require $\tau_{q\neq s} \gg \tau_s$ such that the overall volume $\ccV$ is large, measured in units of $\ell_s$ in the 10d Einstein frame. In this limit the scalar potential is given by \cite{Balasubramanian:2005zx} (see also \appref{F-term scalar potential})
\begin{equation}
\label{F-term potential}
 V_F (\ccV, \tau_s) =  V_{0,F} \left(\frac{\alpha \sqrt{{\tau}_s}e^{-2a_s{\tau}_s}}{c{\ccV}} - \frac{\beta |W_0| {\tau}_s e^{-a_s {\tau}_s}}{{\ccV}^2}+\frac{ \xi\gamma |W_0|^2}{g_s^{3/2}{\ccV}^3}\right)
\end{equation}
with $\alpha, \beta$ and $\gamma$ some positive constants which depend only on $|A_s|$ and $V_{0,F}$ some overall $g_s$-dependent prefactor. Their precise form is given in \eqref{constants of F-term potential}. This potential is already minimised in the axionic (i.e.\ $b_s$) direction. Since it is only the absolute value of $W_0$ and $A_s$ which enter \eqref{F-term potential}, in the following we will write $W_0$ instead of $|W_0|$ etc.

Extremisation with respect to $\tau_s$ in the limit $a_s \tau_s \gg 1$ gives
\begin{equation}\label{Minimising tau_s}
 a_s \tau_s = \log\left(\frac{2\alpha\ccV}{\beta c W_0}\right) - \frac{1}{2} \log \tau_s   =    \log\left(\frac{  4 a_s A_s}{3 c} \frac{ {\cal V}}{W_0}  \right) - \frac{1}{2} \log \tau_s    
\end{equation}
and thus
\begin{equation}
\label{F-term Without tau}
 V_F (\ccV) \simeq V_{0,F} \left(\frac{\xi \gamma W_0^2}{g_s^{3/2} \ccV^3} - \frac{c\beta^2 W_0^2 }{4\alpha \ccV^3 a_s^{3/2}} \log^{3/2}\left(\frac{2 \alpha\ccV }{c\beta W_0}\right)\right),
\end{equation}
where we have neglected terms $\sim \log \tau_s$.
Both terms in the above expression roughly scale like $\sim W_0^2 \ccV^{-3}$. Generically, the same will be true for the value of the $F$-term potential at its minimum. As this minimum is AdS, we need some extra positive contribution to the energy density which lifts the potential at least to Minkowski. In our setup it seems most natural to do this via a $D$-term. As we will see momentarily, such a $D$-term scales like $\ccV^{-2}$ and will thus, in general, give rise to a runaway potential for $\ccV$, unless the size of the above $F$-terms is enhanced. The latter enhancement can be achieved via a large $W_0$. Considering only $\ccV$ and $W_0$, we expect from these arguments that roughly $W_0^2 \sim \ccV$.
\par

Before addressing in detail this question of a dynamical runaway in the closed string moduli space, we first explain why potential instabilities in the open string moduli can generally be avoided geometrically:
During inflation the uplifting $D$-term is due to both the relative gauge flux ${\cal F}_-$ and the overall flux ${\cal F}_+$ on the two D7-branes. 
As detailed in \cite{Hebecker:2011hk}, the end of inflation is marked by a generalised recombination process between the two D7-branes:  ${\cal F}_-$ is responsible for a tachyonic mode in the spectrum between both branes. The  resulting condensation leads to a bound state between the two branes in which the relative $U(1)_-$ is higgsed. 
The remaining bound state continues to carry gauge flux ${\cal F}_+$, whose $D$-term $\xi_+$ is responsible for the uplift to Minkowski/de Sitter after reheating.
To guarantee stability of this $D$-term apart from the potential runaway in the K\"ahler moduli discussed below it must be ensured that no further condensation process occurs.
The only such process would be a generalised recombination between the brane bound state and its orientifold image along their common locus or possibly a recombination between the bound state and a different brane stack in the model. The appearance of a tachyon depends on the pullback of ${\cal F}_+$ to the respective intersection loci and can thus be controlled by a suitable choice of flux, see \cite{Hebecker:2011hk} for details. In particular this requires an explicit choice of orientifold projection from which the brane-image brane intersection can be deduced. While we do not present such a concrete geometry in this work, these arguments are sufficient to show that  a run-away in the open string sector is in general not a problem.
 
\subsection{A Two-Modulus Fluxbrane Inflation Model}
\label{2Model}

\begin{figure}
 \begin{center} 
  \includegraphics[width=0.5\textwidth]{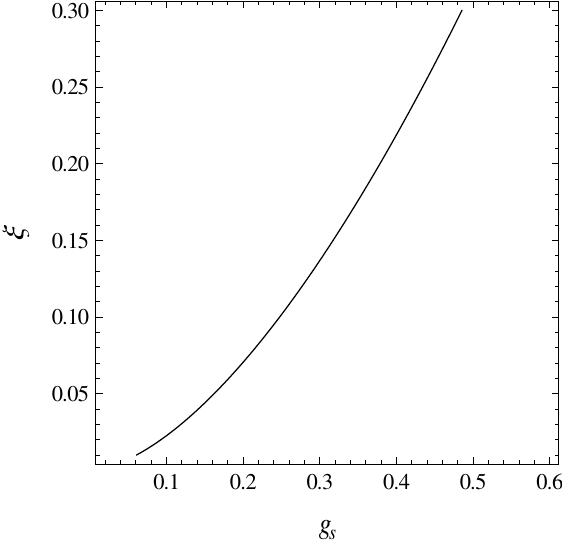} 
 \end{center} 
 \caption{\small Allowed values for $g_s$ and $\xi$ in the simple two-K\"ahler moduli model.}\label{gsxiPlaneSimple} 
 \end{figure}

As an illustrative example consider a two-modulus swiss-cheese model similar to the one discussed e.g.\ in the original LVS publication \cite{Balasubramanian:2005zx}. In such a model the overall volume can be expressed in terms of the two four-cycle volumes $\tau_b$ and $\tau_s$ as
\begin{equation}
 \ccV =  b\tau_b^{3/2} -  c\tau_s^{3/2}
\end{equation}
where $b = \frac{2^{3/2}}{3 ! \sqrt{\kappa_{bbb}}}$, $c = \frac{2^{3/2}}{3 ! \sqrt{\kappa_{sss}}}$, and $\tau_b \gg \tau_s$. Wrapping the fluxed D7-branes around the large four-cycle $D_b$ and choosing flux $\ccF_{\pm} = n_{\pm} [D_b]$ for the overall/relative $U(1)$ theory $U(1)_{\pm}$ of the brane pair induces a $D$-term potential \cite{Hebecker:2011hk,Jockers:2004yj,Jockers:2005zy}\footnote{Since the flux of the relative $U(1)$ theory will annihilate upon brane recombination, we cannot use it for uplifting the minimum value of the potential to zero. Instead, we use $V_D^+$ for the Minkowski uplift, while $V_D^-$ is some additional energy density which is present during inflation and which decays into standard model d.o.f.\ upon reheating.}
\begin{equation}\label{D-termIsotropic}
 V_D^{\pm} (\ccV) = \frac{1}{16\pi \ccV^2}\frac{\left(\int_{\rm D7}J\wedge \ccF_{\pm}\right)^2}{\frac{1}{2}\int_{\rm D7} J\wedge J} = \frac{1}{16\pi \ccV^2}2 n_{\pm}^2 \kappa_{bbb} .
\end{equation}
The full scalar potential thus reads
\begin{equation}\label{FullScalarPotentialSimple}
 V (\ccV) \simeq \frac{V_{0,F} }{\ccV^3} \left(\frac{\xi \gamma W_0^2}{g_s^{3/2}} - \frac{c\beta^2 W_0^2}{4\alpha  a_s^{3/2}} \log^{3/2}\left(\frac{2 \alpha\ccV }{c\beta W_0}\right) + \frac{2n^2 \kappa_{bbb}}{g_s }\ccV\right)
\end{equation}
where $n^2 = n_+^2 + n_-^2 $. Note that from now on we work in a gauge where $e^{K_{\text{cs}}} = 1$. Let $f(\ccV)$ denote the term in the brackets on the right hand side of \eqref{FullScalarPotentialSimple}. Then, in the Minkowski minimum after annihilation of ${\cal F}_{-}$ ($V_D (\ccV)= V_D^+ (\ccV)$) we have $f(\ccV_{\text{min.}}) = f'(\ccV_{\text{min.}}) = 0$. 
Vanishing of $f(\ccV)$ in the minimum yields (to leading order in $a_s \tau_s \simeq \log \left(\frac{4 a_s A_s}{3c}\frac{\ccV_{\text{min.}}}{W_0}\right)$, using also $f'(\ccV_{\text{min.}}) = 0$)
\begin{equation}\label{ccVoverW0}
 \frac{a_s}{g_s}\left(\frac{\xi}{2c}\right)^{2/3} \simeq \log \left(\frac{4 a_s A_s}{3c}\frac{\ccV_{\text{min.}}}{W_0}\right)
\end{equation}
which can be used to rewrite $f'(\ccV_{\text{min.}}) = 0$ as
\begin{equation}\label{W0equation2}
 W_0 \simeq \frac{2}{3}\frac{n_+^2 \kappa_{bbb}}{A_s g_s} \frac{e^{a_s \tau_s}}{\tau_s^{1/2}} .
\end{equation}
Plugging this back into \eqref{ccVoverW0} gives
\begin{equation}
 \ccV_{\text{min.}} \simeq \frac{c \kappa_{bbb}}{2a_s A_s^2} \frac{n_+^2 }{g_s}\frac{e^{2 a_s \tau_s}}{\tau_s^{1/2}} .\label{vmin}
\end{equation}
Setting $n_+ = 5$,\footnote{\label{lowerboundn_+}It turns out that there is a lower bound on $n_+$ which is easy to understand: As $n_-$ is integrally quantised, for a given $n_+$ the uplift to de Sitter cannot be arbitrarily small. However, a large extra $D$-term from the relative $U(1)$ on top of the uplift to Minkowski may potentially wash out the de Sitter minimum for the volume modulus. Therefore, the relative change in the size of the $D$-term before and after inflation cannot be too large or, in other words, $n_+$ has a lower bound. An analytical estimate of this lower bound, performed in \appref{lowerboundn}, gives $n_+ \ge 4$. For what follows we use a slightly more conservative $n_+ = 5$. It is then possible to show numerically that one can indeed obtain a Minkowski minimum for $n_- = 0 $ which is uplifted to a stable de Sitter minimum for $n_- =1$.} $\kappa_{bbb} = 5$, $\kappa_{sss} = 1$ (such that $c = \sqrt{2}/3$),\footnote{Large intersection numbers tend to exacerbate the problems discussed below. Here and in the next section we take $\kappa_{bbb} = 5$ which is, for example, the triple self-intersection number of the quintic \cite{Candelas:1990rm} and appropriate blow-ups thereof \cite{Blumenhagen:2009up,Blumenhagen:2009yv}. } and $A_s = 1$ we find that for $\ccV_{\text{min.}} = 1.7\times 10^6$ the parameters can be chosen to lie in the phenomenologically viable regime (see figure \ref{gsxiPlaneSimple}): 
A value $\xi = 0.1$ implies $g_s \simeq 0.25$. In view of equation \eqref{W0equation2} this means
\begin{equation}\label{W0simple}
  W_0 \simeq  1\times 10^5 .
\end{equation}
It turns out that there is a tension between such a large $W_0$ and the requirement to cancel the D3 tadpole: The authors of \cite{Denef:2004ze} were able to reformulate the tadpole cancellation condition in a way which makes it obvious that, as long as all $F$-terms for the complex structure moduli vanish in the minimum, $W_0 $ or rather $\sqrt{g_s /2 }\ W_0$ is bounded by $\sqrt{\chi (X_4)/24}$, where $\chi (X_4)$ is the Euler characteristic of the associated F-theory fourfold $X_4$. In our example, $\sqrt{g_s /2 }\ W_0 \simeq 3.6\times 10^4$, which would require $\chi (X_4) \simeq 3.2 \times 10^{10}$. To the best of our knowledge, no fourfold with such a large Euler characteristic is known. Therefore, even before considering the production of cosmic strings, the simple two-K\"ahler moduli model turns out not to work quite generically. One can go ahead and try to choose manifolds with different intersection numbers and tune $A_s$ etc. However, we will not go down this road because, to be on the safe side concerning the cosmic string bound referred to in the introduction, we have to consider models beyond this simple one anyhow (see the discussion in section \ref{pheno2}). Instead, we will show that in a slightly more complicated situation $W_0$ can actually be much smaller, such that the tension described above is absent.

Some further comments are in order, which also apply to the more general setup discussed in \secref{Anisotropic Setup}:

\begin{itemize}

\item As $\tau_s$ is given by equation \eqref{Minimising tau_s} we find that, in view of \eqref{ccVoverW0},
\begin{equation}
 \tau_s \simeq \frac{1}{g_s}\left(\frac{\xi}{2c}\right)^{2/3} .\label{taus}
\end{equation}
 This can also be found using a different method (cf.\ \appref{F-term scalar potential}).

\item Uplifting an AdS vacuum through magnetised D7-branes has been 
discussed in different variants in \cite{Burgess:2003ic,Villadoro:2005yq,
Achucarro:2006zf,Haack:2006cy,Cremades:2007ig,Krippendorf:2009zza,Cicoli:2011yh,Cicoli:2012vw}. Unless one appeals to a partial cancellation using charged fields \cite{Cremades:2007ig,Krippendorf:2009zza,Cicoli:2011yh,Cicoli:2012vw}, the $D$-term potential scales as $1/{\ccV}^2$. Since the $F$-term potential 
scales as $W_0^2/{\ccV}^3$, a successful uplift generically requires $W_0^2\sim\ccV$. This is problematic for the following reason:\footnote{
We 
thank Joseph Conlon and Fernando Quevedo for pointing this out.
} 
Estimating the Kaluza-Klein scale on the basis of $T^6$ with equal radii, 
we have $\mkk=\sqrt{\pi}/{\ccV}^{2/3} \sim 1/{\ccV}^{2/3}$. At the same 
time $m_{3/2}\sim W_0/\ccV$, which should be parametrically smaller to justify the use of a 4d supergravity analysis. However, one finds (with $W_0$ normalised as in \cite{Giddings:2001yu})
\beq
\frac{m_{3/2}}{\mkk}=\frac{\sqrt{g_s}}{4\pi}\cdot\frac{W_0}{{\ccV}^{1/3}}
\sim {\ccV}^{1/6}\,.\label{m32r}
\eeq
This `goes the wrong way' at large $\ccV$ (though it does so very weakly).
In \cite{Cremades:2007ig} it was argued that due to the appearance of a large numerical factor $16\pi^4$ in the denominator of $V_D$ it is possible, for $\ccV \sim 10^3$, to uplift the AdS minimum to a stable de Sitter vacuum with $W_0 = \ccO(1)$. In view of \eqref{FullScalarPotentialSimple} we believe that 
the situation is not quite as simple: The only relative factors of $2\pi$ between $F$- and $D$-term contributions come from the definition of $\xi$. 
They suppress the $F$-term, making the situation naively worse, but can be easily compensated by a large $\chi(X_3)$. However, using the explicit formulae (\ref{W0equation2}), (\ref{vmin}) and (\ref{taus}), we can make equation \eqref{m32r} more precise:
\beq
\frac{m_{3/2}}{\mkk}=\frac{n_+}{3\sqrt{\pi}}\sqrt{\frac{\kappa_{bbb}}
{c\sqrt{\tau_s}}}\cdot {\ccV}^{1/6}\,.\label{m32rp}
\eeq
Assuming $n_+ \sim c \sim  \kappa_{bbb} \sim \tau_s \sim 1$ this suggests that, at least in rough numerical agreement with \cite{Cremades:2007ig}, a fairly large ${\cal V}$ can indeed be tolerated in spite of the `parametrical' clash between $m_{3/2}$ and $\mkk$. However, it is not clear that a manifold of swiss-cheese type with such intersection numbers exists. Furthermore, as elucidated above, $n_+ = 1$ does not allow for a stable de Sitter uplift. Larger intersection numbers and a larger value of $n_+$ both deteriorate the situation, reducing the maximal size of the overall volume consistent with the requirement $m_{3/2} < \mkk$. On the other hand, the four-cycle volume $\tau_s$, which could in principle suppress the size of the ratio \eqref{m32rp}, is essentially fixed at a value $\sim 1$ by \eqref{vmin} and the requirement $\ccV_{\text{min.}} = 1.7 \times 10^6$. In particular, with the numbers used and computed in this section we find $m_{3/2}/\mkk \simeq 34$ which means that there is no regime in which the supergravity approximation is valid.

The authors of \cite{Cremades:2007ig,Cicoli:2012fh} furthermore propose to use warping 
to suppress the $D$-term even further. While this is certainly a very appealing possibility, we are hesitant to use it for the $D$-term driving inflation: We 
fear that it might clash with the shift-symmetry that we need to keep our inflaton potential flat. On the other hand, including a further sector of D7-branes with flux in a warped region is certainly an option: It might be used for the uplift from AdS to Minkowski.

Concerning the `inflationary $D$-term', our suggested solution to the `$D$-term-suppres\-sion problem' is a hierarchy between large four-cycles. Given that we have 
several of those with significantly different volumes, we can arrange for
the $D$-term to be parametrically smaller than 1/${\ccV}^2$. 

In other proposals \cite{Cremades:2007ig,Krippendorf:2009zza,Cicoli:2011yh,Cicoli:2012vw} the stabilisation mechanism crucially depends on the presence of non-trivial vevs for some of the charged matter fields which appear in the $D$-term. 
These would arise from the intersection of the mobile D7-brane with other branes in the compactification. A suitable choice of gauge fluxes can in general ensure the absence of such matter fields. Indeed this conforms with our assumptions described at the end of section \ref{The LVS} concerning absence of extra instabilities in the open string sector.

Finally, in \cite{Achucarro:2006zf} the authors consider only one K\"ahler modulus which is charged under the anomalous $U(1)$ and which also appears in the non-perturbative superpotential.

Other proposals for uplifting mechanisms put forward in the recent literature include \cite{Rummel:2011cd,
Cicoli:2012fh}.

\item We will find that in the hierarchical case discussed in \secref{Anisotropic Setup} the $D$-term is not suitable for uplifting the AdS minimum to Minkowski. Therefore, we will need to do the uplift by means of a different mechanism. For concreteness we will consider $\overline{\text{D3}}$-branes.

One might expect that such an $\overline{\text{D3}}$-uplift would also help in the isotropic setup discussed here, since the large $n_+^2$ in \eqref{W0equation2} would be absent. However, this turns out not to be the case: The lack of parametric control over the ratio $m_{3/2}/\mkk$ is still an issue. The size of $W_0$ is reduced only by a factor of $\sqrt{3/5}$ which appears because of the different volume-scaling of the $\overline{\text{D3}}$-brane energy density as compared to the $D$-term. This factor is not nearly sufficient for solving the gravitino mass and the D3-tadpole problems.

\item From $f'(\ccV) = 0$ we find that the $D$-term contribution to the scalar potential (i.e.\ the third term in \eqref{FullScalarPotentialSimple}) is suppressed by a factor of $a_s \tau_s=\frac{a_s}{g_s} \left(\frac{\xi}{2c}
\right)^{2/3}$ relative to the first and second term in that expression. This means that the required uplift (i.e. the value of the $F$-term potential at its AdS minimum) is smaller than the naive parametric expectation, in agreement with the alternative derivation in \appref{F-term scalar potential}. While this tends to exacerbate the `$D$-term-suppression problem' discussed earlier, the effect is already included in equation \eqref{m32rp} and does not change the moderately optimistic conclusion drawn above. 

\item It should be clear from the above that in our scenario SUSY is broken 
at a high scale, $m_{3/2} \sim 10^{-3}$, avoiding the Kallosh-Linde problem \cite{Kallosh:2004yh} in a `trivial' way. While it is interesting to investigate the possibility that, after reheating, a different moduli stabilisation mechanism takes over and low-scale SUSY is recovered \cite{Conlon:2008cj,Antusch:2011wu}, we do not pursue this idea in the present paper.
\end{itemize}

\section{Moduli Stabilisation - Hierarchical Setup}
\label{Anisotropic Setup}
While we saw in the previous section that, within the Large Volume Scenario, it is possible to stabilise the K\"ahler moduli in an AdS minimum at exponentially large overall volume, we ran into trouble trying to uplift the minimum to inflationary dS via a $D$-term potential: For $\ccV \simeq 1.7\times 10^6$ the required size of $W_0$ is in tension with the D3-tadpole constraint and makes $m_{3/2}$ unacceptably large.
On the other hand, this clash is not expected to be a generic feature because in situations with more than two K\"ahler moduli there are further potentially small or large numbers to be considered. These are, in particular, the relative sizes of four-cycles or two-cycles, respectively, and they may well improve the situation, depending on the precise intersection structure.
\par

In fact, considering these more involved models has turned out to be essential for a completely unrelated reason: A more detailed analysis of the phenomenological requirements of fluxbrane inflation reveals that considering isotropic compactification manifolds is actually not enough. In fact, one of the promising outcomes of \cite{Hebecker:2011hk} was that in fluxbrane inflation the energy density of cosmic strings, which are formed upon brane recombination, can be controlled by the relative size of two four-cycles.
\par

We start this section by quickly recalling the most important phenomenological impacts of cosmic strings in brane inflation and motivate the need for a hierarchy.
This discussion is followed by a brief review of how to stabilise the directions transverse to the overall volume in models with more than two K\"ahler moduli via string loop corrections. We then perform a parametrical analysis of the model under consideration and pin down the parametric regime in which the model works. Finally, we perform an explicit calculation and exemplify the situation with a set of concrete numbers for all the quantities that are involved.

\subsection{Cosmic Strings and the Need for a Hierarchy}
\label{pheno2}
At the end of brane inflation cosmic strings are formed generically \cite{Jones:2002cv,Majumdar:2002hy,Sarangi:2002yt}. As these objects carry some energy density they will leave an imprint on the CMB which one should be able to measure in principle. The fact that measurements have not revealed the presence of cosmic strings yet constrains the energy density of these objects. Due to the complicated nature of the bound state formed at the end of inflation, it is actually not immediately clear whether the produced cosmic strings are topologically stable (local) in our fluxbrane scenario. Since a detailed investigation of this interesting question is beyond the scope of this paper, we assume a worst case scenario of local cosmic strings.\footnote{In 
the semilocal case \cite{Vachaspati:1991dz,Hindmarsh:1991jq,Urrestilla:2004eh,Binetruy:2004hho,Dasgupta:2004dw}, the constraints are weakened \cite{Urrestilla:2007sf}.
}
 The resulting constraint can then be phrased as an upper bound on the value of the $D$-term $\xi_-$ during inflation ($\xi_- \le \xi_{\rm crit.}$) \cite{Dvali:2003zh}, which reads
\begin{equation}
 \frac{\left(\int_{\text{ D7}} J\wedge \ccF_- \right)^2}{\frac{1}{2}\int_{\text{ D7}} J\wedge J} \lesssim 8\pi^2 \frac{\alpha}{V_0} \xi_{\rm crit.}^2  .
\end{equation}
We use the results from \cite{Urrestilla:2011gr}, which constrain the product $G \mu$ of the cosmic string tension  $\mu$ and Newton's constant $G$ as $(G \mu)_{\rm crit.} = \frac{1}{4} \xi_{\rm crit.} \simeq 0.42 \times 10^{-6}$, i.e.\
\begin{equation}\label{cosmic string constraint}
 \frac{\left(\int_{\text{ D7}} J\wedge \ccF_- \right)^2}{\frac{1}{2}\int_{\text{ D7}} J\wedge J} \lesssim 9.4 \times 10^{-2} .
\end{equation}
It is thus clear that our compactification manifold needs to have at least two `large' four-cycles with hierarchically different volumes in order for the cosmic string bound to be satisfied. This leads us to consider hierarchical compactification proposals similar to those discussed for example in \cite{Cicoli:2008va,Cicoli:2011yy}. As it turns out\footnote{Here the discussion differs from the one in section 3 of the first version of this paper.} the minimal modification of our previous setup with just one additional K\"ahler modulus is not sufficient for satisfying all phenomenological constraints at the same time. Hence, we will focus on a situation with four K\"ahler moduli.
In this paper we are primarily interested in investigating a general mechanism stabilising the moduli in a manner suitable for fluxbrane inflation rather than in constructing
explicit compactification manifolds that furnish concrete realisations of this mechanism. Reserving the search for such concrete geometries, e.g. along the 
lines of \cite{Cicoli:2011it}, to a later stage we therefore content ourselves with making reasonable assumptions about the topology of the compactification space.
With this understanding, let us assume, for definiteness, that the volume  form is of the type
\begin{equation} \label{volform}
 \ccV =  \frac{1}{2} \kappa_{112} (t^1)^2 t^2  +\frac{1}{2} \kappa_{133} t^1 (t^3)^2 +\frac{1}{2} \kappa_{223} (t^2)^2 t^3 + \frac{1}{6}\kappa_{sss} (t^s)^3 .
\end{equation}
For an overview of our conventions see \appref{Definitions and conventions}. As the $(1,1)$-form $\omega_s$ is dual to a four-cycle which is contractible to a point, $t^s$ is negative.

We choose the following brane and flux setup: The pair of D7-branes is wrapped around the four-cycle dual to the $(1,1)$-form $\omega_2$, while the brane flux is given by $\ccF_{\pm} = n_{\pm} \omega_2$. Thus, the induced D3-brane charge ($\sim \int_{\text{D7}} \ccF_{\pm} \wedge \ccF_{\pm}$) vanishes due to $\kappa_{222}=0$. We now consider the limit $\left|t^1\right| \gg \left|t^2 \right|,\left|t^3\right|$.\footnote{In \secref{The Uplift} we will show that it is indeed possible to stabilise the K\"ahler moduli in this regime.} It turns out that for the K\"ahler metric $K_{T_i \overline{T_j}}$ to be positive definite in this limit we need $\kappa_{133} <0$.
\footnote{As we only specify the intersection numbers (\ref{volform}) rather than a concrete geometry, it is not possible to actually compute the Mori cone. In the general spirit of our approach we take $t^i >0$, $i = 1,2,3$, $\kappa_{112}, \kappa_{223} >0$ as part of the assumptions on our toy model.}

It will be convenient to express all quantities in terms of $\tau_s$, $\ccV$ and the quantities
\begin{equation}
 x \equiv \frac{t^3}{t^1},\hspace{0.3cm}y \equiv \frac{t^2}{t^1} .
\end{equation}
For example, the constraints \eqref{amplitude curvature perturbations} and \eqref{cosmic string constraint} can now be rewritten as
\begin{align}\label{PerturbationAmplitude}
  \ccV^{4/3}y^{2/3} &= \frac{\kappa_{112}^{1/3}}{2^{4/3}} \times 4.2\times 10^8  ,\\
\label{xInequality}
  x^2 & \lesssim \frac{\kappa_{112}}{2 n_-^2 \kappa_{223}^2} \times 9.4 \times 10^{-2}  .
\end{align}
As we will learn presently, the regime in which the model works is at small $x$ and $y$.

\subsection{String Loop Corrections}
\label{Anisotropy}
As we saw in section \ref{The LVS} the interplay between $\alpha'$-corrections to the K\"ahler potential and non-perturbative corrections to the superpotential allows for a minimum of the scalar potential with the overall volume $\ccV$ stabilised at an exponentially large value and the small instanton four-cycle stabilised at $a_s \tau_s \sim \log \left(\ccV/|W_0|\right)$. However, for a model with more than two K\"ahler moduli there will be directions transverse to $\ccV$ which remain flat. As was shown in \cite{Cicoli:2008va,Cicoli:2007xp} these transverse directions may be stabilised by string loop corrections to the K\"ahler potential. In toroidal compactifications those corrections are well known \cite{Berg:2005ja,Berg:2005yu,vonGersdorff:2005bf}. Based on this work the authors of \cite{Berg:2007wt} conjectured that on a general Calabi-Yau manifold string loop corrections to the K\"ahler potential take the form
\begin{align}\label{CorrectionsKahler}
 \delta K_{(g_s)} & = \delta K_{(g_s)}^{\rm KK} + \delta K_{(g_s)}^{\rm W} \nonumber\\
  & = \sum_{i = 1}^{h_{1,1}}\frac{C^{\rm KK}_i (U,\overline{U}) (a_{ij} t^j)}{\Re (S) \ccV} + \sum_{i = 1}^{h_{1,1}} \frac{C^{\rm W}_i (U,\overline{U})}{(b_{ij} t^j) \ccV} .
\end{align}
These corrections originate from the exchange of Kaluza-Klein (KK) modes (with respect to a two-cycle $a_{ij}t^j$) between D7-branes and O7-planes, and of winding (W) modes of strings (along a two-cycle $b_{ij}t^j$ on which the D7-branes intersect). In the example of a toroidal compactification with $\ccO(1)$ values of the complex structure the functions $\ccC^{\rm KK, W}$ were calculated to be of the order $10^{-2}$ (see e.g.\ \cite{Berg:2005yu}).
\par

Although the $g_s$-corrections coming from KK-modes are the leading corrections in the K\"ahler potential in terms of the scaling with the K\"ahler moduli, it was found \cite{Cicoli:2007xp} that in the $F$-term potential actually the $\alpha'$-corrections are dominant. This feature is called \textit{extended no-scale structure} and is crucial to ensure the overall consistency of the approach. Furthermore, as the $g_s$-corrections depend not only on the overall volume $\ccV$ but also on the two-cycle moduli $t^i$, it is intuitively clear that these corrections potentially stabilise the flat directions.
\par

Following \cite{Berg:2007wt} we will assume that in our scenario the $g_s$-corrections take the form
\begin{equation}
 \delta K_{(g_s)} =\frac{g_s }{\ccV}\sum_{i=1}^3 C_{i}^{\rm KK} t^i +  \frac{1 }{\ccV}\sum_{i=1}^3\frac{C_{i}^{\rm W}}{t^i} .
\end{equation}
\par

From these terms one can compute the corresponding leading order corrections to the scalar $F$-term potential as (cf.\ \cite{Cicoli:2007xp})
\begin{align}
 \delta V_{(g_s)} &= V_{0,F} \frac{W_0^2}{\ccV^{10/9}}\left\{\frac{g_s^2 \left(C_1^{\rm KK}\right)^2}{2^{1/3}\kappa_{112}^{2/3}}\frac{1}{y^{2/3}}- \left(4\kappa_{112}\right)^{1/3}\left( C_{2}^{\rm W}\frac{1}{y^{2/3}} + C_{3}^{\rm W}\frac{y^{1/3}}{x}\right) + \ldots\right\}\nonumber\\
\label{g_s corrected scalar potential} 
&= V_{0,F} \frac{W_0^2}{\ccV^{10/3}}\left\{ \ccA g_s^2\frac{1}{y^{2/3}}  + \ccB\frac{1}{y^{2/3}} + \ccC\frac{y^{1/3}}{x} + \ldots \right\} 
\end{align}
with
\begin{equation}
 \ccA = \frac{ \left(C_1^{\rm KK}\right)^2}{2^{1/3}\kappa_{112}^{2/3}}>0, \hspace{0.3cm} \ccB = - \left(4\kappa_{112}\right)^{1/3} C_{2}^{\rm W}, \hspace{0.3cm} \ccC =  -\left(4\kappa_{112}\right)^{1/3} C_{3}^{\rm W} .
\end{equation}
The dots in \eqref{g_s corrected scalar potential} denote terms which are suppressed by further powers of $x$ and $y$ in the limit of small $x,y$ as compared to the leading order contributions. Note that the $C_i^{\rm W}$ can have either sign.

\subsection{Stabilising Ratios of Two-Cycles}
\label{Stabilising Ratios of Two-Cycles}
The string loop corrections discussed in the previous section, together with the $D$-term potential, will stabilise the ratios $x$ and $y$. We now analyse the way in which this happens.
In view of our ignorance concerning the prefactors $\ccA$, $\ccB$, and $\ccC$ in \eqref{g_s corrected scalar potential}, we will \textit{assume} in the sequel that $\ccB$ and $\ccC$ are positive and $g_s^2 \ccA \ll \ccB, \ccC$, such that the dominant terms of the $g_s$-corrections at small $x,y$ are
\begin{equation}\label{relevant loop terms}
 \delta V_{(g_s)} =V_{0,F} \frac{W_0^2}{\ccV^{10/3}}\left\{  \ccB\frac{1}{y^{2/3}} + \ccC\frac{y^{1/3}}{x}  \right\} .
\end{equation}
Thus, $y$ will be stabilised at
\begin{equation}\label{ymin}
 y_{\rm min.} = \frac{2\ccB }{\ccC}x.
\end{equation}
Plugging this back into \eqref{relevant loop terms} yields
\begin{equation}
 \delta V_{(g_s)} =V_{0,F} \frac{W_0^2}{\ccV^{10/3}}\frac{\ccD}{x^{2/3}}
\end{equation}
where $\ccD = \left(3\ccB^{1/3}\ccC^{2/3}\right) / 2^{2/3}$. The runaway of $x$ to infinity will be stopped by the $D$-term potential which, in the hierarchical model specified in \secref{pheno2}, is given by \cite{Hebecker:2011hk,Jockers:2004yj,Jockers:2005zy}
\begin{equation}\label{D-term}
 V_D = \frac{1}{16\pi \ccV^2}\frac{\left(\int_{\rm D7}J\wedge \ccF\right)^2}{\frac{1}{2}\int_{\rm D7} J\wedge J} = \frac{\alpha n^2x^2}{16\pi\ccV^2} \left(1+ \ldots\right).
\end{equation}
The dots denote terms which are higher order in $x$ and $y$, $\alpha = \left(2\kappa_{223}^2\right) /\kappa_{112}$, and\footnote{Note that we will choose to have $n_+ \neq 0$, although, as we will find out presently, the uplift to dS cannot be done via a $D$-term. By keeping $n_+$ the stabilisation of $x$ before and after reheating relies on the same mechanism, which simplifies the discussion.} $n = \sqrt{n_+^2 + n_-^2}$. This term in the scalar potential is enhanced by powers of the overall volume $\ccV$ as compared to \eqref{g_s corrected scalar potential}. For some fixed $\ccV$ the $D$-term will drive $x$ to small values, thereby lowering the relative size of the $D$-term potential with respect to the $F$-terms. Regarding the required suppression of the cosmic string energy density, this is precisely the regime where we want to be. Furthermore, it is this feature which allows us to choose $W_0$ much smaller than in the `warm-up' model of \secref{ModStab}. Minimising in the $x$-direction gives
\begin{equation}
 x_{\rm min.} = \left(\frac{1}{3}\frac{g_s W_0^2 \ccD}{\alpha n^2 \ccV^{4/3}}\right)^{3/8}.
\end{equation}
We thus find a flux-dependent energy density given by
\begin{equation}
 V_{\rm flux}(\ccV) = \frac{g_s W_0^2 }{16\pi \ccV^3}\left(\frac{4}{3^{3/4}}\left(\frac{\alpha n^2 }{g_s W_0^2}\right)^{1/4}\ccD^{3/4}\right) .
\end{equation}
Comparing this expression to \eqref{F-term Without tau} it is apparent that the flux induced energy density is not suitable for an uplift to dS. The reason is that the volume scales similarly to the first term in \eqref{F-term Without tau}. Thus, the same argument which shows that \eqref{F-term Without tau} gives rise to an AdS minimum applies here. Therefore, we need some additional contribution to the vacuum energy density which is suitable for uplifting the minimum to dS. $\overline{\text{D3}}$-branes in a warped throat are a prime candidate for this purpose \cite{Kachru:2002gs,Kachru:2003aw,Kachru:2003sx}.\footnote{We are aware of the recent  discussion \cite{Bena:2009xk,Bena:2011wh,Bena:2012bk} (see also \cite{McGuirk:2012sb}) of potential problems with the supergravity solution corresponding to an $\overline{\text{D3}}$-brane in a warped throat. While these investigations have to be taken very seriously,
we think it is fair to say that, until now, no definite conclusion disproving the viability of such an uplift has been established. As already stated at the end of \secref{ModStab}, D7-branes with flux in a warped region might be a good alternative.} Including their contribution, the full scalar potential reads
\begin{equation}\label{FullScalarPotentialComp}
 V (\ccV) = \frac{g_s W_0^2 }{16\pi \ccV^3} \left(\frac{3}{4}\frac{\xi }{ g_s^{3/2}} - \frac{3}{2}\frac{c }{ a_s^{3/2}} \log^{3/2}\left(\frac{4 a_s A_s }{3c }\frac{\ccV}{W_0}\right)+ \ccE \ccV^{5/3} + \frac{4}{3^{3/4}}\left(\frac{\alpha n^2 }{g_s W_0^2}\right)^{1/4}\ccD^{3/4} \right) .
\end{equation}
The quantity $\ccE$ scales with the fourth power of the warp-factor at the position of the $\overline{\text{D3}}$-brane inside the warped throat. We will assume that it can be tuned arbitrarily by tuning the fluxes which determine the strength of the warping.

\subsection{Parametric Analysis}
\label{Parametric Analysis}
In this section we would like to give an argument why, parametrically, the situation in the hierarchical setup is improved as compared to the `warm-up' model.
\par

Recall that in the model discussed in \secref{ModStab} the tree-level superpotential had to be tuned large, $W_0 \sim \sqrt{\ccV}$, such that the $F$-terms and the $D$-terms had about the same size. The necessity to go in this parameter regime can be seen most easily from \eqref{FullScalarPotentialSimple} (all three terms should be of the same order). In \secref{ModStab} the overall volume $\ccV$ was fixed by the requirement that the right amount of curvature perturbations is produced. The so determined $\ccV$ led to a value of $W_0$ which was incompatible with the D3-tadpole cancellation constraint.
\par

On the other hand, the situation in the hierarchical model looks quite different: The $D$-term features a suppression proportional to $x^2$ and thus the tuning $W_0\sim \sqrt{\ccV}$ is not necessary anymore. The gravitino mass, measured in units of the Kaluza-Klein scale, is given by\footnote{This estimate is based on the approximation $L_{\text{max.}} \sim \sqrt{t^1}$, where $L_{\text{max.}}$ is the volume of the largest cycle on which KK-states propagate.} $m_{3/2} / \mkk \sim W_0/(\ccV^{1/3}y^{1/6})$. Furthermore, we have $t^2 \sim y t^1 \sim y^{2/3}\ccV^{1/3}$ and $t^3 \sim x t^1 \sim W_0^{3/4} \ccD^{3/8} / (\ccV^{1/6}y^{1/3})$. Now we use \eqref{ymin} together with the constraint $\ccV^{2/3}y^{1/3} = \zeta^{-1}$ (cf.\ \eqref{PerturbationAmplitude}) to find
\begin{align}
 & \frac{m_{3/2}}{\mkk} \sim W_0 \sqrt{\zeta},\nonumber\\
 & t^2 \sim \left(\frac{\ccB^{3/2}}{\ccC}W_0\right)^{1/2},\\
 & t^3 \sim  \left(\frac{\ccC}{\ccB^{1/2}}W_0\right)^{1/2} .\nonumber
\end{align}
Here, $\zeta$ is related to $\tilde \zeta$ defined in \secref{pheno1} via $\zeta = \left((2\kappa_{112})^{1/3}N\right)^{-1/2} \tilde \zeta$ where $N$ is the number of e-foldings which we took to be $60$.
We see immediately that, for $\ccB, \ccC = \ccO(1)$, we have to choose $W_0$ somewhat large in order to have $t^2,t^3 \gg 1$. Considering in addition the required smallness of $m_{3/2} / \mkk$ we should choose $1 \ll W_0 \ll \zeta^{-1/2}$. For example, setting $W_0 \sim \zeta^{-1/3}$ we find
\begin{equation}
 t^2 \sim t^3 \sim \zeta^{-1/6} \gg 1, \hspace{0.3cm} \frac{m_{3/2}}{\mkk} \sim \zeta^{1/6} \ll 1 .
\end{equation}
Therefore, in the hierarchical setup it is indeed possible to make $m_{3/2}/\mkk$ parametrically small and, at the same time, have the two-cycle volumes $t^2$ and $t^3$ parametrically large. In the following section we will demonstrate, using explicit numbers, that the tuning \mbox{$W_0 \sim \zeta^{-1/3}$} can be done in a way in which the D3-tadpole constraint is not violated.

\subsection{Quantitative Results in the Hierarchical Setup}
\label{The Uplift}
\begin{figure}
 \begin{center} 
  \includegraphics[width=0.8\textwidth]{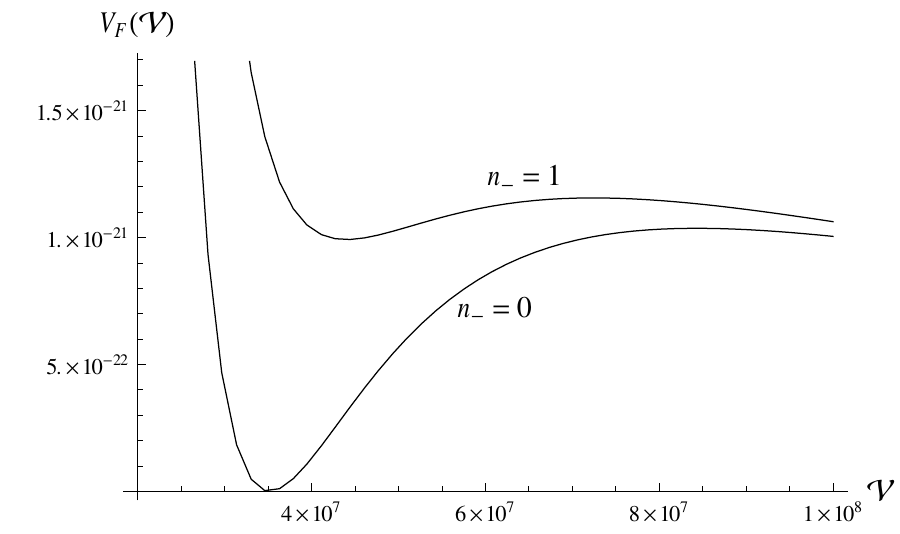} 
 \end{center} 
 \caption{\small Plot of \eqref{FullScalarPotentialComp} for $n_+ =1$, $n_- \in \{0,1\}$}
\label{MinimumAnalysisWithNComplicated} 
\end{figure}
We conclude this chapter by showing that it is possible to consistently choose or compute explicit numbers for all the quantities which are involved in the setup under discussion. This can be done in a way such that all phenomenological constraints are satisfied.
\par

Starting point is the expression \eqref{FullScalarPotentialComp} with $n_- = 0$. Assuming that the threefold has an Euler characteristic\footnote{If, alternatively, we start from a more typical value of e.g. $\chi(X_3) = 100$, the value of $g_s$ increases to $g_s = 0.2$ with all other quantities changed only marginally.} $\chi(X_3) = 5$ we get $\xi \simeq 1.2\times 10^{-2}$. We furthermore choose $W_0 = 2 \times 10^3$, $g_s = 3 \times 10^{-2}$, $\ccD =7 \times 10^{-1}$, $\kappa_{112} = \kappa_{223} = 5$, $\kappa_{sss} = 1$ and thus $c = \sqrt{2}/3$, $A_s = 1$, and $n_+ = 1$. Then, via $V(\ccV_{\text{min.}})= V'(\ccV_{\text{min.}}) = 0$ we can determine $\tau_s \simeq 2.01$. This gives an overall volume $\ccV \simeq 3.5 \times 10^7 $ and, furthermore, $y \simeq 3.9\times 10^{-3} $, $t^2 \simeq 6.0 $. On the other hand, the ratio $x$ is now determined to be $x \simeq 3.3\times 10^{-3}$ and thus $t^3 \simeq 5.1$. Note that this value of $x$ is easily compatible with the cosmic string bound \eqref{xInequality}.
Furthermore, the requirement $V'(\ccV_{\text{min.}}) = 0$ can be used to compute $\ccE \simeq 3.8\times 10^{-14}$.

A plot of the potential \eqref{FullScalarPotentialComp} for $n_- \in \{0,1\}$ is shown in figure \ref{MinimumAnalysisWithNComplicated}. Note that the value $W_0 = 2 \times 10^3$ requires $\chi (X_4) \gtrsim 1.4\times 10^6$ (cf.\ the discussion at the end of \secref{ModStab}). Fourfolds with such Euler characteristics are known in the literature (see e.g.\ \cite{Klemm:1996ts}). Furthermore, in view of \eqref{m32r} we find $m_{3/2}/ \mkk \simeq 0.4$.
Clearly, $m_{3/2}/\mkk$ is only marginally smaller than one and some of our cycle volumes are only marginally larger than the string length. However, as we have demonstrated in \secref{Parametric Analysis}, these crucial inequalities hold parametrically, where the small parameter can be chosen, for example, as the quantity $\zeta$. Then, we of course have to plug in actual numbers, hoping that our parametrically controlled approximation continues to hold in the physical regime. This works sufficiently well to expect that models with the desired type of stabilisation exist.

\section{Flat Directions for the Inflaton}
\label{FlatDir}
In this section we give an overview over possible inflaton mass corrections from the $F$-term scalar potential. While, for most cases, we indicate ways in which these mass terms can be absent or suppressed, we stress that large parts of the subject are still work in progress and will be discussed more thoroughly in a further publication \cite{wip}.

The most direct way in which the D7-brane modulus $\zeta$ can enter the $F$-term potential is through a direct appearance in the tree-level superpotential \cite{Witten:1992fb,Aganagic:2000gs,Lust:2005bd,Jockers:2005zy,Martucci:2006ij,Martucci:2007ey}
\begin{equation}\label{BraneSuperpotential}
 W_{\text{brane}} = \int_{\ccC_5 } \Omega \wedge \tilde{\ccF} .
\end{equation}
Here, $\ccC_5$ is a five-chain ending on the brane divisor $\Sigma$, $\Omega$ is the holomorphic $(3,0)$-form pulled back to $\ccC_5$, and $\tilde{\ccF}$ is the brane flux $\cal F$ continued to the five-chain. For $\tilde{\ccF}|_{\Sigma} \in H^{(1,1)}(\Sigma)$ the wedge product in \eqref{BraneSuperpotential} hence vanishes. Thus it is sufficient to choose $\tilde{\ccF}$ such that $\tilde{\ccF}|_{\Sigma}$ is in the image of $H^2 (X_3)$ under pullback to $\Sigma$: Since for a Calabi-Yau $H^2 (X_3) = H^{1,1} (X_3)$, such fluxes will always give identically vanishing $W_{\text{brane}}$. The flux we consider in this paper is of that kind because it is such flux that generates a D-term potential.

For completeness we furthermore recall from the discussion after equation \eqref{superpotential corrected} that non-perturbative effects from fluxed D3-brane instantons can introduce an explicit inflation dependence of the superpotential, which can be avoided by suitable  constraints on the geometry of the instanton divisors.

Nevertheless, even if we can avoid a direct appearance of the brane modulus $\zeta$ in $W$, the fact that the tree-level superpotential for the complex structure moduli is stabilised at some $W_0 \neq 0$ can lead to a large inflaton mass: Assuming for simplicity a minimal K\"ahler potential $k(\zeta,\overline{\zeta}) = \zeta \overline{\zeta}$ we find that roughly
\begin{equation}
 m_{\zeta} \simeq m_{3/2}
\end{equation}
(cf.\ the discussion at the end of \secref{ModStab}), which is much larger than the Hubble scale and thus spoils slow-roll inflation generically.

The situation is different if the $\zeta$-moduli space possesses a shift-symmetry such that, for example, $k(\zeta,\overline{\zeta}) \equiv k(\zeta + \overline{\zeta})$, i.e.\ the K\"ahler potential is independent of the imaginary part of $\zeta $, which thus remains a flat direction. The potential role of shift symmetries protecting the inflaton mass from dangerous $F$-term contributions  was also discussed previously in the context of D3/$\overline{\text{D3}}$ and D3/D7 inflation (i.e.\ for mobile D3-branes) in \cite{Shandera:2004zy,Hsu:2003cy,Hsu:2004hi,McAllister:2005mq}. However, in these cases one faces some concerns: Usually, in these models inflation proceeds as the D3-brane moves in the radial direction of a warped deformed conifold. The K\"ahler potential of the conifold, however, possesses no radial shift symmetry. More generally, as noted in \cite{McAllister:2005mq}, isometries of the moduli space of D3-positions are not generically present in Type IIB compactifications. In fact, the moduli space of D3-positions, which is nothing but the compactification manifold itself, cannot exhibit a shift symmetry if it has the full $SU(3)$ holonomy. Therefore, only in special examples with manifolds of reduced holonomy, such as compactifications on $K3\times T^2 / \mathbb{Z}_2$, one can hope to find a K\"ahler potential with the desired feature. On the other hand, fluxbrane inflation is in a different position: Even in flat space the leading order potential is flat enough to easily give rise to an inflationary epoch which lasts $60$ e-foldings. Furthermore, the moduli space of D7-positions is not the compactification manifold but rather some other K\"ahler manifold, the K\"ahler potential of which may well exhibit shift symmetries also in more general cases. Therefore, while fluxbrane inflation is a scenario in which one can actually make use of the shift symmetries in the above examples of toroidal and $K3 \times T^2 $ orientifolds, one can even hope to find suitable compactifications beyond these simple models.

In fact, even if the moduli space of D7-brane positions possesses no shift-symmetry generically, there can still be regions in parameter space 
where an approximate shift-symmetry exists. Here we summarise our present understanding within the ongoing investigation \cite{wip}: Some of the 
D7-brane moduli correspond, on the mirror-dual type-IIA orientifold with 
D6-branes, to Wilson-line moduli. Up to instanton and loop-corrections, the
latter enjoy a shift-symmetry originating in the gauge-symmetry of the 
D6-world-volume gauge theory. We can hence expect this shift symmetry to
be present in our setting if, in addition to being at large volume, we also
insist in being near the large-complex-structure point. We finally note 
the possible interest in shift-symmetries of this type in the context of 
Higgs-physics with a high-scale SUSY breaking \cite{Hebecker:2012qp,Ibanez:2012zg,Chatzistavrakidis:2012bb}.

It is interesting to consider the global properties of the $\zeta$-moduli space. As the moduli space of the axio-dilaton $S$ is non-trivially fibered over the former, the K\"ahler potential has the structure
\begin{equation}\label{KahlerPotential}
 K \supset - \log (S + \overline{S} - k(\zeta,\overline{\zeta})) .
\end{equation}
Typically, the moduli space of the D7-brane modulus $\zeta$ is covered by a set of coordinate patches with appropriate transition functions. Generically, the K\"ahler potential $k(\zeta,\overline{\zeta})$ on the D7 moduli space is not globally defined but rather undergoes a transformation $k(\zeta, \overline{\zeta}) = k'(\zeta', \overline{\zeta'}) + f(\zeta') + \overline{f(\zeta')}$ for a transition function $\zeta = \zeta(\zeta')$. For the full K\"ahler potential $K$ in \eqref{KahlerPotential} to remain invariant, this transformation has to be absorbed by a simultaneous redefinition of the axio-dilaton: $S = S' + f(\zeta')$. Invariance of $K$ implies invariance of the superpotential: $W(S) = W'(S',\zeta')$. In other words, $W$ can not be independent of the brane position moduli in all patches. As $W$ is holomorphic there is then no chance of having a manifest shift-symmetry in all patches. However, all we need is to find a flux choice for which the D7-brane coordinate does not appear in $W$ in one particular coordinate patch. After a change of coordinates it will be some combination of the axio-dilaton $S$ and the brane modulus $\zeta$ which is a flat direction in the scalar potential. In \cite{wip} we will demonstrate this feature in the example of a $K3\times T^2/\mathbb{Z}_2$ compactification in detail. The ambiguity of having $W(S)$ or $W(S, \zeta)$, depending on the coordinate patch, is related to the ambiguity in the definition of brane or bulk fluxes. This, in turn, has to do with the $SL(2, \mathbb{Z})$ monodromy affecting $S$ and $(F_3,H_3)$ at 7-brane positions \cite{Denef:2004ze}.
Pertinent investigations of the 7-brane superpotential include the recent \cite{Alim:2009bx,Grimm:2009ef,Grimm:2009sy,Jockers:2009ti,Grimm:2010gk,Alim:2011rp,Xu:2012ks}.

The analog of this issue in the case of inflation with mobile D3-branes has been discussed in detail in the literature \cite{McAllister:2005mq,Burgess:2006cb}. The superpotential depends non-perturbatively on the volume modulus (which, in this setup, plays a role analogous to $S$ in the case of fluxbrane inflation). The gauge kinetic function, which enters this non-perturbative term, receives one-loop corrections which depend on the D3-brane position. For a toroidal compactification these corrections were analysed in \cite{Burgess:2006cb} and shown to respect a discrete shift-symmetry, reflecting the compactness of the torus. No continuous shift-symmetry is preserved by these corrections. Crucially, there is no mechanism by which the appearance of the D3-brane coordinate in the superpotential can be avoided.

Finally, we note that the string loop corrections, which were used in \secref{Anisotropy} to stabilise the relative size of the two large four-cycles, generically depend on open string moduli. The precise form of this dependence in toroidal models can in principle be extracted from \cite{Berg:2005ja,Berg:2005yu}. The relevance of these corrections for the flatness of the inflaton potential is presently under investigation \cite{wip}.

\section{Consistency of the Effective Theory}
\label{consistency}
While the question of a potential inconsistency of (constant) FI terms in supergravity is not a novel issue 
(see, e.g.~\cite{Witten:1985bz,Binetruy:2004hho}), it has attracted an 
increased amount of interest more recently \cite{Komargodski:2009pc,
Dienes:2009td,Komargodski:2010rb,Seiberg:2010qd,Distler:2010zg,Banks:2010zn,
Hellerman:2010fv}. Given that $D$-term inflation in its original form
\cite{Binetruy:1996xj,Halyo:1996pp} relies on the presence of a (constant) 
FI term and that the existence of consistent gravity models with this feature 
is doubtful, we find it necessary to devote a section of our paper to this 
issue. For example, the arguments above have led the authors of 
\cite{Gwyn:2011tf} to conclude that D3/D7 inflation as well as fluxbrane 
inflation are subject to rather stringent constraints. As we will explain, 
we believe that our construction can not come into conflict even with 
the most stringent no-go theorems concerning FI terms that are being debated. 

The viability of $D$-term inflation in view of supergravity constraints on 
FI terms has also been discussed in \cite{Binetruy:2004hho}. However, since 
our perspective and (part of) our conclusions are different, we believe that 
it is worthwhile to revisit this issue.

\subsection[Issues in String $D$-Terms]{Issues in String {\boldmath $D$}-Terms}
Recall that the $D$-term in supergravity is given in general by 
\cite{Bagger:1990qh}
\begin{equation}\label{D-termKillingVector}
 \xi = i K_i X^i (z)\,,
\end{equation}
where $K$ is the K\"ahler potential and $X(z)$ is the holomorphic Killing 
vector generating the (gauged) isometry of the moduli space. We denote the 
coordinates $z^i$ on that moduli space collectively by $z$. The $D$-term 
potential then reads
\begin{equation}\label{D-termSUGRA}
 V_D = \frac{g_{\text{YM}}^2}{2} \xi^2\,.
\end{equation}
\par

The consistency question alluded to above is, roughly speaking, under which 
circumstances one may write 
\begin{equation}
 \xi = i K_i X^i (z) + \xi_0
\end{equation}
for some constant $\xi_0 \neq 0$. For our purposes, the precise answer to this
question is, in fact, irrelevant. We are only interested in string-derived 
models and hence for us it is sufficient to know that no such constant arises 
in the low-energy limit of string compactifications \cite{Dine:1987xk,Atick:1987gy,Douglas:1996sw,Ibanez:1998qp} 
(at least there are no such 
examples). Moreover, as we will work out in more detail momentarily, our 
$D$-term potential is described by the (undebated) part $iK_i X^i (z)$. Since 
this has given us a viable model of inflation, one might think that the 
`FI term-issue' in fluxbrane inflation is thus closed. 

Things are not quite as simple, though. Given that the $D$-term 
potential drives inflation, the moduli in $i K_i X^i (z)$ must be 
stabilised. In fact, this was the main theme of the present investigation. 
Hence one might expect to encounter, somewhere between the 
moduli-stabilisation scale and the SUSY-breaking scale, an effective theory 
with constant FI term. This would not only be potentially inconsistent, it 
turns out to be technically impossible in models where no FI term is 
originally present \cite{Binetruy:2004hho,Komargodski:2009pc,
Komargodski:2010rb}. How can any stringy version of $D$-term inflation then 
exist? The answer suggested in \cite{Binetruy:2004hho} was to have a {\it 
small} $F$-term potential giving a {\it large} mass to the relevant moduli, 
which might be possible with a special choice of K\"ahler potential. Jumping 
ahead, our answer is different: In our scenario the SUSY breaking scale is enhanced as compared to the scale at which the K\"ahler moduli are stabilised. This fact is easy to understand: As we will confirm momentarily, the K\"ahler moduli naturally have masses $m_{\tau}^2  \sim V_D \sim \ccV^{-2}$. On the other hand the gravitino mass is given by $m_{3/2}^2 \sim W_0^2 / \ccV^2$. Recall that, as a result of the approximate no-scale structure in the $F$-term potential and the requirement $V_D\sim V_F$, we work at parametrically large $W_0$. Therefore, the gravitino mass is parametrically larger than  $m_{\tau}$. Due to this particular hierarchy of scales our model avoids the above constraints `trivially'. We will come back to this fact at the end of this section.

\subsection{Moduli Masses in Fluxbrane Moduli Stabilisation}
\label{Moduli masses}

We first put our $D$-term potential in the standard $N=1$ supergravity 
form following \cite{Haack:2006cy}: Let $D_j$ be a divisor with dual 2-form 
$[D_j]$ and let the fluxed D7-brane be wrapped on a divisor $D_{\ccF}$. 
The four-cycle modulus $\tau_j$ parametrising the size of $D_j$ gets charged 
under the $U(1)$ on the D7-brane if the flux living on the intersection 
$D_j \cap D_{\ccF}$ is non-vanishing. Since the symmetry which is gauged is 
an axionic shift symmetry, the corresponding Killing vector is just $X_j = 
i q_j$ with $q_j$ the charge of $\tau_j$. Thus, in the low-energy 
effective action a $D$-term
\begin{equation}
\label{Dterm-shift}
 \xi = - q_j K_{\tau_j}
\end{equation}
appears, where $K$ is the K\"ahler potential. The charge $q_j$ depends on 
the flux \cite{Haack:2006cy}: $q_j \sim \int_{D_{\ccF}} [D_j] \wedge \ccF$. In 
particular, one can show that with this input that \eqref{Dterm-shift} is 
equivalent to \cite{Jockers:2004yj,Jockers:2005zy}
\begin{equation}\label{D-termGeneral}
 \xi = \frac{1}{4\pi}\frac{\int_{D_{\ccF}}J\wedge \ccF}{\ccV}\,,
\end{equation}
which was used in \secref{ModStab} and \secref{Anisotropic Setup}.

In the simple two-K\"ahler moduli example at the end of \secref{ModStab}, we 
wrapped the D-brane on $D_b$ and chose a flux $\ccF = n [D_b]$. As 
$\kappa_{bbb}$ is non-zero, $\tau_b$ is charged under the $U(1)$, generating 
a $D$-term of the form \eqref{D-termGeneral}. The potential terms relevant 
for the mass of $\tau_b$ are \eqref{F-term Without tau} 
and \eqref{D-termIsotropic}. The corresponding K\"ahler potential is
\begin{equation}
 K = -3 \log \tau_b + \ldots
\end{equation}
For simplicity we compute the mass in the Minkowski minimum (the result 
changes only by an $\ccO(1)$ factor when going to de Sitter). Working in 
addition to leading order in $\log\left(\frac{2\alpha \ccV}{c \beta W_0}
\right)$, we find
\begin{equation}\label{VolumeModulusMass}
 m_{\tau_b}^2 = \left.\frac{1}{K_{\tau_b \tau_b}}\partial^2_{\tau_b}V
\right|_{\tau_b^{\text{min.}}} = \left.\frac{3}{4} V_D \right|_{\tau_b^{\text{min.}}},
\end{equation}
where $V_D$ is given in \eqref{D-termIsotropic}.
\par

This can be compared to the mass of the vector boson which gauges the 
axionic shift-symmetry. The relevant terms in the Lagrangian are
\begin{equation}
\ccL \supset - \frac{1}{4g_{\text{YM}}^2} F^2 + K_{i\overline{ \jmath}} 
D_\mu z^i \overline{D^\mu z^j},
\end{equation}
with
\begin{equation}
 D_\mu z^i = \partial_\mu z^i - A_\mu X^i (z) .
\end{equation}
Thus, using also \eqref{D-termKillingVector} and \eqref{D-termSUGRA} the 
gauge boson mass is
\begin{equation}
m_V^2 = 2 g_{\text{YM}}^2 K_{T_b\overline{ T_b}} \left|X^{T_b}\right|^2 = 
\left.\frac{4}{3}V_D \right|_{\tau_b^{\text{min.}}} .
\end{equation}
We conclude that the masses of vector boson and corresponding K\"ahler 
modulus are of the same order of magnitude. Moreover, since $V_D\sim H^2$,
both masses are related to the Hubble scale. While the purist might object 
both to calling this $D$-term inflation (since $V_F\sim V_D$) and to calling
it single-field inflation (since $m_{\tau_b}\sim H$), we are, for the time 
being, satisfied with this outcome. 

Finally, the analysis in \appref{lowerboundn} gives
\begin{equation}
 \frac{\delta V_D}{V_D} \lesssim \frac{2}{3^3}
\end{equation}
where $\delta V_D$ represents the additional energy density due to $\ccF_- 
\neq 0$ during inflation. This implies that the Hubble scale during inflation 
is given by $H_{\text{infl.}} \lesssim 0.07 \times V_D $. Therefore, $\tau_b $ is 
actually somewhat heavier than the parametric analysis above suggests 
($m_{\tau_b} / H_{\text{infl.}} \gg 1$) and its dynamics can be disregarded during 
inflation.

We note that our result can be understood more generally (see e.g.~\cite{
Binetruy:2004hho,Komargodski:2009pc}): In unbroken SUSY the mass of the 
vector and the mass of the volume modulus are the same because the $U(1)$ is 
higgsed by the axionic scalar from the volume superfield (in our case 
$T_b$). This equality can only be lifted by SUSY breaking. Thus, if the 
mass of the volume modulus is stabilised at a scale much above the vector 
mass, supersymmetry must be broken at this high scale. As result, there can 
be no energy domain where $\tau_b$ is consistently integrated out while the 
gauge boson is kept as a dynamical degree of freedom in a supersymmetric 
theory. In other words, as mentioned earlier, even an {\it effectively} 
constant FI term can not arise. 

In our specific setting (at least in the toy model version of \secref{ModStab}),
we have $H^2 \sim m_V^2 \sim m_{\tau_b}^2 \sim V_D \sim V_F $,
as demonstrated above. By contrast, $m_{3/2}^2$ is much larger. This is
due to the (approximate) no-scale cancellation which makes $V_F$ smaller
than its naive parametrical size $|e^K W_0^2|$. Hence, we are indeed more
than safe from any regime with unbroken SUSY and an effectively constant FI term. Of course, the analysis in the present section dealt just with the toy model of \secref{ModStab}. An analogous discussion of the hierarchical model of \secref{Anisotropic Setup} is qualitatively similar but much more involved. While the various `low-lying' mass scales from $H$ to $m_V$ are now somewhat different, the much larger size of $m_{3/2}$ is a generic feature. It will continue to ensure that SUSY is broken before the moduli are frozen.

\section{Conclusions}
We have studied moduli stabilisation in fluxbrane inflation. In this scenario, the role of the inflaton is played by the relative position of two D7-branes, attracted towards each other by non-supersymmetric gauge flux. This can be viewed as a variant of $D$-term inflation where, as is well-known, a very small FI term is required to reproduce the observed magnitude of CMB fluctuations. In our context, this implies large brane volume and hence, in general, large compactification volume ($\sim 10^7$ in string units).

We therefore work in the Large Volume Scenario, where the interplay of $\alpha'$- and instanton corrections stabilises an exponentially large overall volume. The resulting non-SUSY AdS-vacuum is then uplifted by a $D$-term to realise the inflationary almost-de-Sitter phase. Stability requires the $F$-term to be roughly of the same size as the $D$-term, which we ensure by using a parametrically large flux-potential $W_0$. This entails two problems: a dangerously large (close to the KK-scale) gravitino mass and an enormous D3 tadpole (in excess of the largest known fourfold Euler numbers). Moreover, as is common in $D$-term inflation in general, the large cosmic string scale is problematic.

Fortunately, a rather natural generalisation of the simplest construction can resolve all of the above issues: We explicitly include two additional K\"ahler moduli (the minimum in large volume models being two). In this case, three of the K\"ahler moduli are stabilised at relatively large values. Their relative size is fixed by the interplay of loop corrections and the $D$-term. This introduces two new parameters -- the ratios of two-cycle volumes -- which can take small values ($\sim 1/300$) in a concrete model. In appropriate settings, one of these smallish numbers suppresses the $D$-term (thereby lowering the required value of $W_0$) as well as the cosmic string tension. Thus, while we do not provide an explicit Calabi-Yau orientifold and brane configuration, we are able to demonstrate the phenomenological viability of our scenario with reasonable assumptions concerning topological data and loop-correction coefficients. 

To be more specific, in \secref{Parametric Analysis} we demonstrated that we have parametric control in the regime of $\zeta \ll 1$, where $\zeta$ is related to the amplitude of density perturbations. Therefore, all approximations made in the stabilisation procedure depend on this physical observable. If one matches the COBE value, the effective field theory is marginally under control, whereas the approximations become better and better as the amplitude of density perturbations falls below the observed value.

In the hierarchical model discussed in \secref{Anisotropic Setup} it was not possible to achieve the uplift of the AdS vacuum to a Minkowski vacuum via a flux contribution to the $D$-term: The phenomenological requirements, which need to be imposed on the model, were too restrictive for that purpose. While, for the present analysis, we contented ourselves with the idea of realising the uplift with $\overline{\text{D3}}$-branes instead, it would be interesting to further investigate the $D$-term uplifting proposal in more general setups.

To realise our flat inflaton potential, we had to assume a choice 
of 3-form flux by which the relevant D7-brane modulus is not stabilised. Nevertheless, a generic K\"ahler potential for the brane moduli leads,
in presence of non-zero $F$-terms, to the familiar $\eta$-problem. Referring 
to a forthcoming publication for details, we propose to overcome this 
problem as follows: By mirror symmetry, some of the D7 position moduli 
correspond to D6 Wilson lines. The latter enjoy a shift-symmetric K\"ahler potential at large Type IIA volume. Hence, we expect a shift-symmetric 
inflationary K\"ahler potential in Type IIB at large complex structure (in addition to large volume). It is far from obvious to which extent this 
idea will survive the loop corrections which we also use. This is presently
under investigation.

Finally, we commented on certain consistency issues in the context of the recent `supergravity FI term debate'. We have no proper (`constant') FI term but rather a conventional $D$-term potential (or `field-dependent FI term'). The latter comes from the gauging of an isometry of the K\"ahler moduli space. We show explicitly that, in our construction, the vector boson mass, the K\"ahler modulus mass and the Hubble scale are parametrically of the same order of magnitude. Hence, the potential problem of an `effectively constant FI term' at some intermediate energy scale (which would hint at some hidden inconsistency) does not arise. Thus, $D$-term inflation (at least in our definition, i.e. allowing for comparable stabilising $F$-terms) is well and alive independently of the existence of constant FI terms. 

So far, we can only view our investigation as a small step in the ongoing 
struggle to eventually establish inflation in string theory (or rule it out). For the future of our proposal, much will depend on our ability to understand how D7-brane moduli enter the K\"ahler potential and superpotential at subleading order.

\section*{Acknowledgements}
We thank Max Arends, Joseph Conlon, Konrad Heimpel, Benjamin Jurke, Christoph Mayrhofer, Fernando Quevedo and Christoph Schick for helpful discussions and comments. Furthermore, we gratefully acknowledge the referee's criticism of an earlier version of this paper.
AH and TW gratefully acknowledge the hospitality of the Isaac Newton Institute, Cambridge. This work was supported by the Transregio TR33 ``The Dark Universe'', by the "Innovationsfonds FRONTIER" in Heidelberg and partially by the Cluster of Excellence ``Origin and Structure of the Universe'' in Munich.

\appendix

\section*{Appendix}

\section{Definitions and Conventions}
\label{Definitions and conventions}
In this appendix we collect definitions and conventions used in this article.
\par

The string length is given by $\ell_s=2\pi\sqrt{\alpha'}$. We take the transformation from string to 10d Einstein frame to be
\begin{equation}
	g^S_{MN}=e^{\frac{\phi}{2}}g^{\text{E}10}_{MN}
\end{equation}
and we use $g_s=\left< e^{\phi} \right>$, $\tau=e^{-\phi}+iC_0$.\footnote{Note that the relation between $\tau$ defined here and the supergravity variable $S$ used e.g.\ in \eqref{ksz} is non-trivial \cite{Jockers:2004yj,Jockers:2005zy}.} The volume $\ccV$ of the compactification space is defined by
\begin{equation}
	\ccV = \frac{1}{\ell_s^6}\int{d^6 x \sqrt{g_6^{\text{E}10}}} .
\end{equation}
The transformation from 10d Einstein frame to 4d Einstein frame is given by
\begin{equation}
	g_{\mu\nu}^{\text{E}10}=\frac{1}{ \ccV}g_{\mu\nu}^{\text{E}4} .
\end{equation}
The four-dimensional Planck mass is then given by
\begin{equation}
	M_p^2=\frac{4\pi}{\ell_s^2}.
\end{equation}
It is set to one in all 4d-field-theory formulae.
The K\"ahler form $J$ in the Einstein frame is expanded in a basis of $(1,1)$-forms $ J=  t^i\omega_i$, such that the volume of the manifold can be written as
\begin{equation}
	\ccV=\frac{1}{6}\int{ J\wedge  J \wedge  J}=\frac{1}{6}\kappa_{ijk}  t^i t^j t^k.
\end{equation}
The volumes of the four-cycles dual to the $(1,1)$-forms $\omega_i$ are defined via $\tau_i = \partial \ccV / \partial t^i$.

\section{{\boldmath $F$}-Term Scalar Potential}
\label{F-term scalar potential}
The main purpose of this appendix is to analyse the $F$-term potential discussed in \secref{The LVS}.
Such a potential arises in the original Large Volume Scenario as proposed in \cite{Balasubramanian:2005zx} as well as in more elaborate versions thereof \cite{Cicoli:2008va}, which is the case of interest for us.

Starting point is the expression\footnote{There seems to be a disagreement in the literature concerning the overall prefactor of the supergravity potential (see \cite{Giddings:2001yu} and \cite{Conlon:2005ki}). However, this factor is irrelevant for our purposes as we can simply choose to work with a differently normalised $W_0$.}
\begin{align}\label{V_F start}
 V_F =& \frac{e^K}{8\pi} \left[K^{s\overline{s}} a_s^2 |A_s|^2 e^{- 2 a_s \tau_s} \right.\nonumber\\
& - a_s K^{s\overline{p}}\partial_{\overline{p}}Ke^{-a_s \tau_s}\left\{W\overline{A}_s e^{i a_s b_s }+ \overline{W} A_s e^{-ia_s b_s}\right\}\\
& + \left. \frac{3\xi |W_0|^2 }{4g_s^{3/2}\ccV}\right]\nonumber
\end{align}
which is obtained after plugging \eqref{superpotential corrected} and \eqref{Kahler corrected} into the standard supergravity formula for the $F$-term potential
\begin{equation}
 V = \frac{e^{K}}{8\pi}  \left(K^{a \overline{b}} D_{a} W D_{\overline{b}} \overline{W} - 3|W|^2\right),
\end{equation}
expanding in leading order in $1/\ccV$, and neglecting all terms $\propto e^{-a_{p}\tau_{p}}$, $p\neq s$ (cf.\ \cite{Balasubramanian:2004uy}).
\par

Consider the second line of equation \eqref{V_F start}. We can rewrite the term in the brackets as
\begin{equation}\label{aux eq 1}
 2 |W_0| |A_s |  \cos \left(\arg (W_0) - \arg (A_s) + a_s b_s \right) .
\end{equation}
Furthermore, using the identity
\begin{equation}
 K^{s\overline{p}}\partial_{\overline{p}}K = -2\tau_s + \text{higher orders in $1/\ccV$}
\end{equation}
(cf.\ e.g.\ \cite{Cicoli:2007xp}) it is clear that minimising $V_F$ with respect to the axion $b_s$ will give \mbox{$\cos(\ldots) \rightarrow -1$} in \eqref{aux eq 1} and thus the second term in \eqref{V_F start} becomes $-4a_s \tau_s e^{-a_s \tau_s} |W_0| |A_s|$.
\par

Now we turn to the first line in \eqref{V_F start}: Using $\ccV(\tau_p) = \tilde{\ccV}(\tau_{p\neq s}) - c \tau_s^{3/2}$ we find
\begin{equation}
 K_{s \overline{s}} \simeq\frac{3}{8} \frac{c}{\ccV \tau_s^{1/2}} \ , \ \ K_{p \overline{s}} \simeq -\frac{3}{4}  \frac{c(\partial_{p} \ccV)\tau_s^{1/2}}{\ccV^2} ,
\end{equation}
i.e.\ $K_{p\overline{q}}$ is block-diagonal in leading order in $1/\ccV$. Therefore, $K^{s\overline{s}} \simeq \frac{8}{3}\frac{\ccV \tau_s^{1/2}}{c}$ in leading order. Combining all the results we find
\begin{equation}
\label{Full F-term potential}
 V_F = V_{0,F}\left(  \frac{\alpha \sqrt{\tau_s}e^{-2a_s\tau_s}}{c\ccV}  -\frac{ \beta|W_0| \tau_s e^{-a_s \tau_s}  }{\ccV^2} + \frac{\gamma\xi |W_0|^2 }{g_s^{3/2}\ccV^3} \right) 
\end{equation}
with
\begin{equation}\label{constants of F-term potential}
 V_{0,F} = \frac{ g_s e^{K_{\rm cs}}}{16\pi }, \ \ \alpha = \frac{8 a_s^2 |A_s|^2}{3} , \ \ \beta = 4 a_s  |A_s|, \ \ \gamma = \frac{3}{4} .
\end{equation}

We now compute the large volume minimum of \eqref{Full F-term potential}. To this end, we evaluate
\begin{equation}
	\frac{\partial V}{\partial \ccV} = 0 , \ \ \frac{\partial V}{\partial \tau_s} = 0.
\end{equation}
The first condition gives
\begin{equation}
	\ccV = \frac{ \beta |W_0| c \sqrt{\tau_s} e^{a_s \tau_s}}{\alpha}\left(1 - \sqrt{1-\frac{3\alpha \gamma \xi}{g_s^{3/2} c \beta^2 }\frac{1}{\tau_s^{3/2}}}\right) 
	\label{eq:VolumeSolution}
\end{equation}
while the second equation reads
\begin{equation}
	\frac{\ccV \alpha e^{-a_s \tau_s}}{\beta c |W_0|\sqrt{\tau_s}} = \frac{1-a_s \tau_s}{\frac{1}{2} - 2a_s \tau_s} .
\label{eq:tauSolution}
\end{equation}
This can be rearranged using \eqref{eq:VolumeSolution}:
\begin{equation}
	1 - \sqrt{1-\frac{3\alpha \gamma \xi}{g_s^{3/2} c \beta^2 }\frac{1}{\tau_s^{3/2}}} =  \frac{1-a_s \tau_s}{\frac{1}{2} - 2a_s \tau_s} .
	\label{eq:TauCondition}
\end{equation}
For $a_s \tau_s \gg 1$, this equation simplifies and we obtain to leading order
\begin{align}
	\tau_s &= \frac{1}{g_s}\left( \frac{4 \gamma \alpha \xi}{c\beta^2} \right)^{2/3} = \frac{1}{g_s}\left(\frac{\xi}{2c}\right)^{2/3} ,
	\label{eq:TauValueAppendix}\\
	\ccV &= \frac{ \beta |W_0| c \sqrt{\tau_s} e^{a_s \tau_s}}{2\alpha} = \frac{ \beta |W_0| c }{2\alpha g_s^{1/2}} \left( \frac{\xi}{2c} \right)^{1/3} e^{\frac{a_s}{g_s} \left( \frac{\xi}{2c} \right)^{2/3}}.
	\label{eq:VolumeValueAppendix}
\end{align}
\par

From the above analysis and the definition of $\xi$ below equation \eqref{Kahler corrected} it is clear that the requirement $a_s \tau_s \gg 1$ is true as long as
\mbox{$-\chi(X_3) \gg \frac{4c}{\zeta(3)} (2\pi g_s)^{3/2}$}. Since $\chi(X_3)=2(h^{(1,1)}-h^{(2,1)})$ and $h^{(1,1)} = 2$, this sets a lower bound on the number of complex structure moduli which is, however, rather easy to satisfy for all perturbative values of $g_s$. For instance, the manifold $\mathbf{P}^4_{[1,1,1,6,9]}$ of \cite{Balasubramanian:2005zx}, on which the large volume scenario was first constructed explicitly, has 272 complex structure moduli.

To compute the value of the potential $V_F$ at the minimum, we solve \eqref{eq:tauSolution} for $\ccV$ and plug it into \eqref{Full F-term potential} to obtain
\begin{equation}
	V_F = V_{0,F} \frac{\alpha^2}{\beta c^2 |W_0|}e^{-3a_s\tau_s} \left(  X -  X^2 + \frac{\alpha\gamma\xi}{\beta^2 c g_s^{3/2}}\tau_s^{-3/2} X^3 \right),
\end{equation}
where
\begin{equation}
	X = \left(\frac{\frac{1}{2}-2a_s\tau_s}{1-a_s\tau_s}\right).
\end{equation}
Rewriting \eqref{eq:TauCondition} yields
\begin{equation}
	\frac{\alpha\gamma\xi}{\beta^2 c g_s^{3/2}}\tau_s^{-3/2} = \frac{1}{3}\left( 2 X^{-1} - X^{-2}\right)
\end{equation}
and therefore
\begin{equation}
	V_F = V_{0,F} \frac{\alpha^2}{\beta c^2 |W_0|}e^{-3a_s\tau_s} \left( \frac{2}{3} X - \frac{1}{3}  X^2 \right) \approx V_{0,F}  \frac{\alpha^2}{\beta c^2 |W_0|}e^{-3a_s\tau_s} \left( - \frac{1}{a_s\tau_s} \right)
\end{equation}
for $a_s\tau_s \gg 1$. Remarkably, the minimum value of the $F$-term potential is suppressed by a factor of $a_s\tau_s$ relative to its natural value. Using \eqref{eq:TauValueAppendix} and \eqref{eq:VolumeValueAppendix} we finally obtain
\begin{equation}\label{F-term minimum Appendix}
 V_F = -  \frac{3 M_p^4 \sqrt{g_s} e^{\ccK_{cs}} }{128\pi^2}\frac{c |W_0|^2}{ \ccV^3}\left(\frac{\xi}{2c}\right)^{1/3}
\end{equation}
where we chose to explicitly write $\ccV$ and $|W_0|$ in order to get a feeling for the size of the $F$-term potential in its minimum. It is clear, that via \eqref{eq:TauValueAppendix} and \eqref{eq:VolumeValueAppendix} one can express the minimum value solely in terms of $\xi$, $g_s$, $c$ etc.

\section{Lower Bound on the Brane Flux Quanta}
\label{lowerboundn}
In this appendix we give a rough estimate for the lower bound on $n_+$ as motivated in \secref{2Model}. To this end we write \eqref{FullScalarPotentialSimple} as
\begin{equation}
 \frac{V(\ccV)}{V_{0,F}} = \frac{1}{\ccV^3}f(\ccV)
\end{equation}
where $f(\ccV) = A - B\log^{3/2}(C\ccV) + D\ccV$. Furthermore, we choose to expand the potential at the minimum $\ccV_{\text{min.}}$ as
\begin{equation}\label{PotentialExpansion}
 \frac{V(\ccV)}{V_{0,F}} = \frac{f''(\ccV_{\text{min.}})}{2 \ccV^3}(\ccV-\ccV_{\text{min.}})^2 + \ldots
\end{equation}
Maximising \eqref{PotentialExpansion} gives\footnote{Actually, as $(\ccV_{\text{max.}} - \ccV_{\text{min.}}) > \ccV_{\text{min.}}$ the expansion \eqref{PotentialExpansion} breaks down. However, the calculation still gives a first idea for the required size of $n_+$ which can be confirmed numerically.} $\ccV_{\text{max.}} = 3 \ccV_{\text{min.}}$ and
\begin{equation}
 \frac{V(\ccV_{\text{max.}})}{V_{0,F}} \simeq \frac{2}{3^3} \frac{f''(\ccV_{\text{min.}})}{\ccV_{\text{min.}}} .
\end{equation}
Computing $f''(\ccV_{\text{min.}})$, keeping only the leading term in $\log (C\ccV_{\text{min.}})$, and using $f'(\ccV_{\text{min.}}) = 0$ one finds
\begin{equation}
 \frac{V(\ccV_{\text{max.}})}{V_{0,F}} \simeq \frac{2  }{3^3} \frac{D}{ \ccV^{2}_{\text{min.}}} .
\end{equation}
An estimate of how big the uplift can be such that it does not destroy the local minimum of the potential is given by requiring
\begin{equation}
 \frac{V(\ccV_{\text{max.}})}{V_{0,F}} \gtrsim \frac{\delta V(\ccV_{\text{min.}})}{V_{0,F}} = \frac{\delta D}{\ccV_{\text{min.}}^{2}} .
\end{equation}
This then implies
\begin{equation}
 \frac{\delta D }{D} = \frac{n_-^2}{n_+^2}\lesssim \frac{2  }{3^3}  .
\end{equation}
Thus, $n_+ \ge 4$, $n_- = 1$ is ok. Obviously, this is only a very coarse analysis. However, the result can be confirmed very easily by a straightforward numerical analysis.

\bibliography{ref-inflation}  
\bibliographystyle{utphys}
\end{document}